\documentstyle[aps,epsf,eqsecnum]{revtex}
\begin{document}
\preprint{
DOE/ER/40762-104\hspace{1cm}
U.of MD PP\#97-045}
\title{Baryon resonances in a chiral confining model.}
\author{Y.~Umino and M.K.~Banerjee}
\address{Department of Physics, University of Maryland, 
College Park, MD 20742--4111, U.S.A.}
\date{\today}
\maketitle
\begin{abstract}
In this two part series a chiral confining model of baryons is used to 
describe low--lying negative parity resonances $N^*$, $\Delta^*$, 
$\Lambda^*$ and $\Sigma^*$ in the mean field approximation. 
A physical baryon in this model consists of interacting valence quarks, 
mesons and a color and chiral singlet hybrid field coexisting inside a 
dynamically generated confining region. 
This first paper presents the quark contribution to the masses and wave 
functions of negative parity baryons calculated with an effective 
spin--isospin dependent instanton induced interaction. It does not include 
meson exchanges between quarks.
The three--quark wave functions are used to calculate meson--excited baryon 
vertex functions to lowest order in meson--quark coupling. 
When the baryons are on mass--shell each of these 
vertex functions is a product of a coupling constant and a form factor.
As examples, quark contributions to $N^*$ hadronic form factors as well
as axial coupling constants are
extracted from the vertex functions and problems with the 
analytical behaviour of the model form factors are discussed. 
The second paper will examine the mesonic
corrections to excited baryon properties in the heavy baryon and one--loop
approximations.

\noindent PACS number(s): 12.40.Yx, 14.20.Gk, 14.20.Jn
\end{abstract}

\vfill\eject
\section{INTRODUCTION}
\label{introduction}
An interesting and sensitive test for models of baryon structure will soon 
be performed at the recently completed Jefferson Lab facility when the 
radiative decay 
widths of the first two excited states of the $\Lambda$ hyperon, 
$\Lambda(1405)$ and 
$\Lambda(1520)$, will be determined directly in a quadrupole coincidence 
measurement 
experiment \cite{CLAS}. There are four such widths to be measured, two for 
each excited 
hyperon radiatively decaying into ground states $\Lambda^0$ and 
$\Sigma^0$, and so far 
none of the predictions of various models of baryons inspired by 
Quantum 
Chromodynamics (QCD) agree with each other \cite{dar83,kax85,umi93,sch95}.
Calculation of these widths is extremely model dependent involving the 
wave functions of excited hyperons as well as transition operators both of 
which must be determined in a self consistent manner. It is therefore a 
challenge for 
{\em any} model of baryon structure to predict correctly all four widths. 
Discrepancies among existing model predictions are large enough to be able 
to distiguish experimentally among various results. But unfortunately, 
presently available 
experimental data on the four radiative decay widths either disagree by an 
order of 
magnitude \cite{mas68,ber84} or have been extracted in a model dependent 
manner 
\cite{bur91}. 

Phenomenologically this test is particularly welcome since almost all the 
models of baryon structure can describe the static properties of the 
ground state 
octet and decuplet baryons rather well, and therefore it is very difficult 
to judge the success of any given model based on its description of the 
ground states alone. 
In addition, it is often the case that a given model which can reproduce 
the properties of 
ground state baryons will run into difficulties when extended to describe 
the    
excited states. An example of this difficulty is the removal of the 
center--of--mass motion in relativistic quark 
models where no rigorous method to project out the spurious degrees of 
freedom exist. 
Thus a successful description of ground state baryons in any model is a 
necessary but not a sufficient condition for the model to be realistic, 
and 
the upcoming direct precision measurements of the excited 
hyperon radiative decay widths at Jefferson Lab present an excellent 
testing ground for any model of baryons under development. 

One such model is the chiral confining model (CCM) of baryons 
\cite{ban93,ban97} 
where a baryon is described as a color singlet composite object consisting 
of interacting valence 
quarks, mesons and a color and chiral singlet hybrid field $\chi$. The 
model produces  
absolute quark confinment using a 
mechanism introduced by Nielsen and P\`{a}tkos \cite{nie82} to dynamically 
generate a confining region, or a ``bag'', inside which the above 
constituents coexist to form a baryon. Although the model has not been 
directly derived from QCD, it is at least consistent with the essential 
properties of the underlying theory of strong interactions such as chiral
invariance and the behaviour in the limit of large number of colors, $N_c$. 

The initial version of the CCM\cite{ren90} revealed two major problems. 
The model prediction for the product of the nucleon mass and the quark rms 
radius was too large, while that for $N - \Delta$ mass splitting was too 
small. McGovern, {\it et al} \cite{McG} showed there was an approximate 
scaling law for the CCM which stated that if either the mass or the 
size of the nucleon was kept fixed at some particular value all other 
predictions of the CCM were also more or less fixed. The problems were  
resolved \cite{kim93} by extending the model to include the 't~Hooft 
interaction \cite{tho76,shi80}. By fitting the $N 
- \Delta$ mass difference 
to determine the strength of the effective 't~Hooft interaction and using 
simple estimates to correct for the center--of--mass motion, the 
prediction for 
the product of the nucleon mass and the quark rms radius were reduced 
to an acceptable value in the range of 3.8 to 4.4, the empirical value 
being about 3.5. 

In this two--part series the CCM is extended to the strangeness sector and 
is used to examine the properties of low--lying negative parity 
$N^*$, $\Delta^*$, $\Lambda^*$ and $\Sigma^*$ baryons in the mean field
approximation. Excited baryons are regarded as RPA excitations of the ground 
state baryons and includes both particle--hole excitations of the valence 
quarks as well as creation of pseudoscalar mesons. 
The first paper considers excited baryons built with valence quarks 
in $S_\frac12$, $P_\frac12$ and $P_\frac32$ states and examines 
the contribution of such states to the masses, wave functions and hadronic 
form factors using a spin--isospin dependent instanton induced interaction 
\cite{tho76,shi80}.
Baryons obtained with only the valence quarks are refered to as 
''bare'' baryons. Mesonic corrections to properties of bare excited 
baryon will be presented in a followinging paper.

The role of gluons in the CCM appears in three distinct ways:
\begin{itemize}
\item[] First, the gluon condensate of the vacuum manifests itself through 
the appearance of the color dielectric function~\cite{nie82}. The 
fluctuation of the 
gluon condensate appears in the form a $0^{++}$ hybrid field~\cite{ban93}. 
\item[] Second, the instanton, a feature of {\it classical} Euclidean QCD, 
appears through the 
$2N_f$ ($N_f$ being the number of flavors) fermion interaction introduced 
by 
't~Hooft~\cite{tho76,shi80} 
to represent approximately the role of instantons on the quark 
contribution to the partition 
function. In the absence of this idea it would not be possible to include 
the role of instantons 
in a mean field theory in the Minkowski world. 
\item[] Third, the role of the exchange of {\it quantum} gluons is described 
through meson exchanges\cite{bron86,bbc87,ball87}. 
\end{itemize}
Note that perturbative one--gluon exchange (OGE) interaction 
cannot be included
in this model as that would lead to double counting. 
Since OGE interaction is widely used in quark models by many authors, 
the differences between three--quark wave functions obtained with the 
instanton induced and OGE interactions are examined.

In order to diagonalize the bare baryon Hamiltonian, $H_{\rm Bare}$, 
it is necessary to remove the spurious components in 
the excited baryon wave functions corresponding 
to the translation of the center--of--mass of the composite baryon. 
For the CCM, where a baryon is a composite object consisting of 
relativistic quarks, mesons and the $\chi$ field, seeking an 
exact solution to 
this problem is an impractical task. Thus in this work a prescription 
originally used in the study of excited baryons in the MIT bag model 
by DeGrand \cite{deg76b} is used to approximately eliminate the 
unphysical states.
The method is able to predict the observed number 
of low--lying negative parity $N^*$ and $\Delta^*$ states but is strictly 
valid only in the non--relativistic and SU(6) spin--flavor limits. 
A simple test of this prescription is presented in the Appendix~B in order to 
estimate the relative amount of 
spurious component present in the excited hyperon wave function for different 
strange quark masses. 
After eliminating the spurious states $H_{\rm Bare}$ is diagonalized by 
assuming equal masses for up, down and strange quarks to be consistent
with the center--of--mass removal prescription. Corrections from flavor 
symmetry breaking to the hyperon masses are given to first order by
modifying the single particle energies of strange quarks after 
diagonalization. 

The three--quark wave functions for excited baryons are 
used to calculate meson--bare excited baryon vertex functions to
lowest order in meson-quark coupling. When the baryons are on
mass--shell each of these vertex functions is a product of a
coupling constant and a form factor.
Meson--bare baryon form factors are extracted from the vertex 
functions and used to determine the quark contribution to the axial 
coupling constants of negative parity nucleons. It is pointed out that these 
hadronic form factors do not have the correct analytical behaviour in the 
complex $q$--plane.
In the second paper these vertex functions are used to determine 
meson cloud contribution to physical baryon properties in the 
heavy baryon and one--loop approximations.

Essential features of the CCM are presented in the following section together
with single particle energies and spinors of confined
quarks in the ground and
relevant excited states. The model Hamiltonian for bare baryons is then
defined by using the 't~Hooft interaction as the residual interaction
between quarks.
In Section~III, this Hamiltonian is diagonalized and bare masses and
wave functions of negative parity baryons $N^*$, $\Delta^*$, $\Lambda^*$
and $\Sigma^*$ are presented and compared to other
model calculations.
These three--quark wave functions are used in Section~IV to calculate
meson--bare excited baryon vertex functions. Corresponding form factors 
for negative parity $N^*$ resonances and their axial coupling constants 
are also presented as examples and problems
with the results discussed. 
Finally, the paper is summarized in Section V. 

\section{THE CHIRAL CONFINING MODEL AND THE 't~Hooft INTERACTION}
\label{model}
\subsection{The Chiral Confining Model of Baryons}
\label{ccm}

The CCM~\cite{ban93,ban97} is based on the color dielectric theory 
proposed by Nielsen and P\`{a}tkos~\cite{nie82}. These authors introduced 
two collective variables, $K(x)$, a color singlet, charge--conjugation 
even object, and $B^a_\mu(x)$, a coarse--grained gluon field expressed in 
terms of averages of link operators for loops contained in a 
hypercube of side $L$. Upon integrating out the QCD gluon fields in favor 
of these new 
collective variables Nielsen and P\`{a}tkos obtained the following Lagrangian 
in the form of a derivative expansion,
\begin{equation} {\cal L}_{NP}=\bar{\psi}(x)
\biggl[ iK(x)\frac12\stackrel{\leftrightarrow}
{\partial\!\!\!/}-K(x)m_q-gB\!\!\!\!/\;(x)\biggr] \psi(x)
-\frac{K(x)^4}{4}G^a_{\mu\nu}G^{a\,\mu\nu}+\ldots.  
\label{NPL1}  
\end{equation}  
In terms of the gauge field, $ B^a_\mu/K$, the coarse grained field tensor is
\begin{equation}  G^a_{\mu\nu}=\partial_\mu \frac{B^a_\nu}{K}-\partial_\nu
\frac{B^a_\mu}{K} + f^{abc}\frac{B^b_\mu}{K}\frac{B^c_\nu}{K}.  
\label{NPL2} 
\end{equation}
From the gluonic term one identifies  $\epsilon=K^4$ as the color 
dielectric function.  Nielsen and P\`{a}tkos conjectured that 
\begin{equation} 
\langle K\rangle_{vac}=0. 
\label{Kvac1} 
\end{equation}

The CCM adds four new features to the Nielsen and P\`{a}tkos model. First,
the conjecture shown in Eq.~(\ref{Kvac1}), crucial for the CCM, was justified 
from the lattice gauge point of view by Lee {\it et al.}~\cite{lee89}.
It is customary to introduce the field $\chi(x)$ as
\begin{equation} 
K(x)=g_\chi\chi(x), 
\label{Kchi}
\end{equation} 
whereupon $g_\chi$ has the 
dimensions of $L$. The second feature in CCM is the demonstration, 
through large $N_c$ analysis~\cite{ban93}, that  
$\chi(x)$ is a hybrid and not a glueball field\footnote{While the properties 
of the QCD vacuum is dominated by the glueball field, changes produced by 
baryon matter are dominated by hybrids.}. This is crucial for proper large 
$N_c$  behavior of CCM itself.

The third feature in CCM is the conjecture 
that there exists a suitable gauge in 
which the coarse--grained gluon field $B^a_\mu(x)$ may be integrated out in 
favor of meson fields.  The meson fields were described with the 
Gell--Mann L`{e}vy linear $\sigma$--model~\cite{GeLe60}. 
This allowed the introduction, 
through $\langle 0\mid \sigma\mid 0\rangle = -93$ MeV, of an ever--present 
mass like term for the quarks in a perfectly chiral invariant manner. 
Thus field variables in 
the CCM are quarks, mesons and the hybrid field $\chi$. The vanishing of 
$K(x)$ in the vacuum means that
\begin{equation}
\langle 0| \chi | 0 \rangle=0. 
\label{chivac}
\end{equation}

The meson--quark coupling may involve a factor $K^n$. 
The fourth feature of CCM is the fixing of the exponent $n$. This was 
found by comparing the $K$--dependence of the four--fermion interaction 
obtained by integrating out the $B_\mu^a$ fields of the Nielsen and  
P\`{a}tkos Lagrangian, Eq.~(\ref{NPL1}), with that obtained by integrating 
out the mesons in the CCM Lagrangian. It was found that $n=-1$. 

Finally, canonical quark fields were introduced using the transformation 
$\sqrt{g_\chi\chi}\psi\rightarrow\psi$. In terms of the canonical quark 
fields the effective CCM Lagrangian, extended to SU(3) flavor group is 
\begin{equation}
{\cal L}_{\rm CCM} =  
\overline{\psi} \Bigl[ 
\frac{i}{2} \stackrel{\leftrightarrow}{\partial\!\!\!/} -\; m_q
+ \frac{ g_\phi }{(g_\chi \chi)^2} 
\biggl(\sigma + i\gamma_5 \lambda^a\phi^a 
+ \cdot\cdot\cdot \biggr) \Bigr] \psi
+ {\cal L}_\chi + {\cal L}_{\rm meson}.
\label{eq:CCM}
\end{equation}
where $\phi^a$ are the pseudoscalar octet mesons with flavor index $a$, 
$(a = 1 \rightarrow 8)$, and the ellipses refer to other mesons such as
$\omega, \rho$ or $A$. The Gell--Mann matrices $\lambda^a$ are 
normalized to ${\rm Tr}\:(\lambda^a \lambda^b) = 2\delta_{ab}$.  
The inclusion of vector mesons requires that CCM Lagrangian use the Lee 
and Nieh's form for ${\cal L}_{\rm meson}$~\cite{LeNi68}.

Two simple forms for the $\chi$ potential, pure mass and quartic, have 
been considered. This paper 
will use only the pure mass form given by 
\begin{equation}
{\cal L}_\chi = \frac{1}{2} \partial_\mu \chi \partial^\mu \chi 
- \frac{1}{2} m_\chi^2 \chi^{2}. 
\label{eq:CHI}
\end{equation}
Extensive SU(2) calculations have shown that the meson fields produced 
by the valence quarks themselves play relatively less important role in 
the baryon structure. Hence a reasonable approximation is to set the 
mesons fields at their vacuum values. Thus 
\begin{equation} 
\langle 0 \mid\sigma\mid 0 \rangle=-F_\pi,
\label{sigvac} 
\end{equation}
and all other meson fields are fixed at the value zero. In addition, the
parameter $g_\phi$ in Eq.~(\ref{eq:CCM}) is set to $g_\pi$. The 
resulting model is called the Toy model. It must be understood that it is 
used solely to obtain reasonable approximations to the valence spinors and 
the $\chi$ field. For example, in order to calculate meson--baryon 
coupling constants the full CCM Lagrangian must be used to obtain the meson 
source currents.

The Toy model Lagrangian is
\begin{equation}
{\cal L}_{\rm Toy} = \overline{\psi} \left( \frac{i}{2} 
\stackrel{\leftrightarrow}{\partial\!\!\!/} -\; m_q
- \frac{ g_\pi F_\pi }{(g_\chi \chi)^2} \right) \psi 
+ {\cal L}_\chi.
\label{Toylag}
\end{equation}
The bag formation mechanism depends on two critical ingredients specified 
by Eqs.~(\ref{chivac}), $\langle 0 \mid\chi\mid 0 \rangle=0$, and 
(\ref{sigvac}), 
$\langle 0 \mid\sigma\mid 0 \rangle\neq 0$ and incorporated in the Toy 
model. The Euler Lagrange equation for the 
quark field from the Toy model is
\begin{equation}
(i\partial\!\!\!/ - m_q)\psi = -\frac{g_\pi F_\pi}{(g_\chi \chi)^2} \psi.
\label{eq:MOTION}
\end{equation}
Thus a quark cannot exist in the region where $\chi=0$. 
However because of the coupling between quarks and the $\chi$ field,  
valence quarks change the value of the $\chi$ field away from its vacuum 
value and make it nonzero. This region is the dynamically created bag 
where a quark can exist.

There are four adjustable parameters in ${\cal L}_{\rm Toy}$,
$m_q$, $m_\chi$, $g_\chi$ and $g_\pi$. The first is left fixed at the SU(3) 
flavor limit value of $m_q = 7.5$ MeV. Corrections to hyperon masses 
from SU(3) breaking are discussed below.
McGovern {\it et al}~\cite{McG} showed that the Toy model possesses 
approximate scaling and, as a consequence, the combination 
$[g_\pi F_\pi m_\chi^2/g_\chi^2]^\frac15$, 
which has the dimensions of $L^{-1}$, is the 
only operative parameter, not the individual values of $ m_\chi, g_\chi$ 
and $g_\pi$. In the description of ground state 
properties of $N$ and $\Delta$ \cite{kim93,kim94}
these parameters are fixed to be $m_\chi$ = 1200 MeV, $g_\chi$ = 0.08 
MeV$^{-1}$ and $g_\pi = 4$. The 't~Hooft interaction 
introduces two additional parameters, the interaction strength 
$C_s$ and the regulator parameter $r_c$, which are also fixed in 
\cite{kim93,kim94} to be $C_s = 1.6\; {\rm fm}^2$ and 
$r_c = 0.25\; {\rm fm}$. 
These values of model parameters are used in the present work.
\subsection{The toy model}
\label{toymodel}
The toy CCM Hamiltonian corresponding to ${\cal L}_{\rm Toy}$ of 
Eq.~(\ref{Toylag}) is given by
\begin{equation}
H_{\rm Toy} = \int\! d^3r\left[\psi^\dagger(\vec{r}\;)\left\{-i 
\vec{\alpha}\cdot\vec{\nabla} + \beta \left( m_q + 
\frac{g_\pi F_\pi}{(g_\chi \chi)^2} \right)\right\}\psi(\vec{r}\;) + 
\frac12\Bigl[(\vec{\nabla}\chi)^2+m_\chi^2\chi^2\Bigr]\right],
\label{eq:TOY}
\end{equation}
and the corresponding energy function when the three valence quarks occupy 
the same spinor state $u(\vec{r}\;)$ is
\begin{equation}
E_{\rm Toy} = \int\! d^3r\; u^\dagger(\vec{r}\;)\left\{-i 
\vec{\alpha}\cdot\vec{\nabla} + \beta \left( m_q + 
\frac{g_\pi F_\pi}{(g_\chi \chi)^2} \right)\right\}u(\vec{r}\;)+E_\chi, 
\label{Etoy}  
\end{equation}
where $E_\chi$ is the {\em classical} $\chi$ field energy given by
\begin{equation}
E_\chi = 
\int\! d^3r\frac12 \Bigl[(\vec{\nabla}\chi)^2+m_\chi^2\chi^2\Bigr].
\end{equation}
In the mean field treatment of the CCM the ground state problem is solved 
by making $E_{\rm Toy}$ of Eq.~(\ref{Etoy}) stationary with respect to the 
quark spinor $u(\vec{r}\;)$ and the $\chi$ field, subject to the normalization 
$\int\! d^3r \; u^\dagger(\vec{r}\;)u(\vec{r}\;)=1$.

Let $\chi^{(0)}$ be the $\chi$ field resulting from the mean field 
treatment of the ground state and $E^{(0)}_\chi$ the corresponding classical 
$\chi$ field energy. Then from Eq.~(\ref{eq:TOY}), the CCM Hamiltonian 
describing non--interacting quarks moving in a fixed mean $\chi$ field is 
\begin{equation}
H^{(0)}_{\rm Toy}=\int\! d^3r\; \psi^\dagger(\vec{r}\;)\left\{-i 
\vec{\alpha}\cdot\vec{\nabla} + \beta \left(  m_q +  
\frac{g_\pi F_\pi}{(g_\chi \chi^{(0)})^2} \right) \right\}\psi(\vec{r}\;) + 
E^{(0)}_\chi.
\label{H0Toy}
\end{equation} 
$H^{(0)}_{\rm Toy}$ describes the foundation of the quark model used in this
work. The lowest energy eigenspinor of this 
Hamiltonian is the one which makes 
the energy function given by Eq.~(\ref{Etoy}) stationary. The numerical 
value of $E^{(0)}_\chi$ is found to be about 300 MeV 
\cite{kim94}. Being the same 
for all baryon states it plays no role in determining the 
spin--flavor compositions of excited baryons. 
In the present quark shell model description of excited baryons, the 
low--lying negative parity states
are constructed by exciting a single quark from the ground $S_{1/2}$ state 
to either the 
$P_{1/2}$ or $P_{3/2}$ state. Adapting the notation $S \equiv S_{1/2}$, $P 
\equiv P_{1/2}$ and $A \equiv 
P_{3/2}$, quark spinors for $S$, $P$ and $A$ states are defined as
\begin{eqnarray}
u_S (\vec{r}\;) & \equiv & 
\left( 
\begin{array}{c}
G_0(r) \\ i\vec{\sigma\;}\cdot\hat{r} F_0(r)
\end{array}
\right) \xi_S(\hat{r}),
\label{eq:SWAVE} \\
u_P (\vec{r}\;) & \equiv & 
\left( 
\begin{array}{c}
G_2(r) \\ i\vec{\sigma\;}\cdot\hat{r} F_2(r)
\end{array}
\right) \xi_P(\hat{r}),
\label{eq:PWAVE} \\
u_A (\vec{r}\;) & \equiv & 
\left( 
\begin{array}{c}
G_3(r) \\ i\vec{\sigma\;}\cdot\hat{r} F_3(r)
\end{array}
\right) \xi_A(\hat{r}).
\label{eq:AWAVE}
\end{eqnarray}
These spinors satisfy the eigenvalue equation
\begin{equation}
H^{(0)}_{\rm Toy}\; u_i(\vec{r}\;) = \epsilon_i\; u_i(\vec{r}\;),
\label{eq:EIGEN}
\end{equation}
where $i =  S, P$ or $A$.  

Figure~1 shows the upper, $G_i(r)$, and the lower, $F_i(r)$, radial 
functions in the quark spinor for $S$, $P$ and $A$ states defined in 
Eqs.~(\ref{eq:SWAVE}) to (\ref{eq:AWAVE}) for $m_u=m_d=7.5$ MeV and 
$m_s=300$ MeV.  
Also shown in Figure~1a is the radial profile for the $\chi^{(0)}$ 
field which appears in $H^{(0)}_{\rm Toy}$. 
The eigenvalues of $S$, $P$ and $A$ spinors of $H^{(0)}_{\rm Toy}$ as 
defined in Eq.~(\ref{eq:EIGEN}) for $m_q = m_{u,d}$ and $m_q=m_s$ are 
presented in Table~\ref{H0energies}. In addition, Table~\ref{stable} 
shows the values of the quark scalar charge ${\cal S}_i$ defined as,
\begin{equation}
{\cal S}_i \equiv \int\!d^3\!r\; \overline{u}_i u_i = 
\int\!dr\; r^2 \left( G^2 - F^2 \right)_i,
\label{eq:SCALAR}
\end{equation}
with the normalization 
\begin{equation}
\int d^3\!r u_i^\dagger u_j=\delta_{ij}.
\label{normspin}
\end{equation}
As expected the higher the current quark mass $m_q$ smaller is the lower 
components of the quark spinors and higher are the scalar charges. Note, 
however, that changes in ${\cal S}_i$ are $40$\% and larger.

In the present work the effect of quark mass difference $m_s-m_{u,d}$ is 
treated perturbatively. Only the spinors with $m_u=m_d=m_s=7.5$ MeV are 
used to evaluated matrix elements. SU(3) breaking effects on hyperon masses 
are estimated 
using the quark scalar charge given in Eq.~(\ref{eq:SCALAR}) as follows.
Three--quark wave functions for excited hyperons are first obtained by 
diagonalizing the mass matrix including the 't~Hooft interaction. 
In general these wave functions are linear 
combinations of states with the excited quark in either the $P$ or the $A$ 
state. After diagonalization, single particle energies for strange quarks 
are modified from $\epsilon_i$ to $\epsilon_i + \delta\epsilon_i$ where 
\begin{equation}
\delta\epsilon_i=(m_s-m_q)\int\! d^3r\; \bar{u}_i u_i. 
\label{delepsi}
\end{equation}
Using the scalar charges given in Table~\ref{stable} for $m_q = 7.5$ MeV, 
the numerical values for $\delta\epsilon$, are
\begin{eqnarray}
\delta\epsilon_S & = & (m_s-m_q) \int\! d^3r\; \bar{u}_S u_S = 184\; 
{\rm MeV}, \\
\delta\epsilon_P & = & (m_s-m_q) \int\! d^3r\; \bar{u}_P u_P = 100\; 
{\rm MeV}, \\
\delta\epsilon_A & = & (m_s-m_q) \int\! d^3r\; \bar{u}_A u_A = 163\; 
{\rm MeV}.
\end{eqnarray}
For example, consider a state where the strange quark is excited 
to the $A$ state while the remaining two non--strange quarks are in their 
ground states. The sum of single quark energies for this state is modified 
from 
$2\epsilon_S + \epsilon_A = 1170$ MeV to $2\epsilon_S + \epsilon_A + 
\delta\epsilon_A = 1333$ MeV where $\epsilon_i$'s are the single quark 
energy eigenvalues obtained with $m_q$ = 7.5 MeV.

\subsection{The 't~Hooft interaction} 
\label{Hooft}
Phenomenological consequences of the 't~Hooft interaction in baryon 
structure have been investigated using various quark models by Kochelev 
\cite{koc85}, Shuryak and Rosner \cite{shu89}, 
Oka and Takeuchi \cite{oka89}, Blask {\em et.al} \cite{bla90}, Kim and 
Banerjee\cite{kim93,kim94}, Klabu\v{c}ar \cite{kla94} and most recently by 
Takeuchi \cite{tak94,tak95}. 
The role of the 't~Hooft interaction in the excited baryon mass spectra has 
previously been examined by Blask {\em et al.} \cite{bla90} and Takeuchi 
\cite{tak94,tak95} 
using variants of the non--relativistic quark model. The model used by 
Blask {\em et al.}  
consists of a linear confining potential simulating a string--like 
confinement and the 't~Hooft interaction which was used as the residual 
interaction without any OGE interactions. Using this model they managed to 
reproduce the 
qualitative features of meson and baryon mass spectra up to masses of 
order 2 GeV including the $\eta - \eta'$ splitting. Shuryak and Rosner 
\cite{shu89} 
also used a similar constituent quark model to parametrize the ground state 
baryon mass spectrum using only the two--body effective instanton 
induced force. 
Thus the constituent quark model seems to give a reasonable description of 
the low--lying baryon and meson mass spectra with either the OGE or 
the 't~Hooft 
interaction as the residual interaction between a pair of quarks.

The Hamiltonian for the constituent quark model 
used by Oka and Takeuchi \cite{oka89} and later by Takeuchi 
\cite{tak94,tak95} 
includes both the OGE and 't~Hooft interactions in the follwoing way,
\begin{equation}
H = K + V_{\rm conf} + (1 - p_{\rm III})V_{\rm OGE} + p_{\rm III} 
V_{\rm 't~Hooft},
\label{eq:TAKEUCHI}
\end{equation}
where $K$ and $V_{\rm conf}$ are quark kinetic energy and confining 
potentials, respectively. The parameter $p_{\rm III}$ is introduced to 
adjust the strength of the OGE potential, $V_{\rm OGE}$, relative to 
that of the 't~Hooft interaction potential, $V_{\rm 't~Hooft}$.\footnote{This 
Hamiltonian treats the classical 
't~Hooft and quantum OGE interactions on an equal footing. In the CCM
gluon exchanges are effectively taken into account by meson exchanges which
in the mean field theory are included as a quantum corrections to the 
classical 't~Hooft interaction.
Thus, assuming a Hamiltonian of this form would lead to overcounting in 
the present work.} This parameter is in the range of 0.3 to 0.4 when the 
model is used to fit the $\eta - \eta'$ mass splitting. Takeuchi used this 
Hamiltonian to calculate the masses of excited nucleons up to the $N=2$ 
harmonic oscillator level and the nucelon--nucleon force in the 
quark cluster model \cite{tak94,tak95}.  

It was realized early in the development of the NRQM by Isgur and 
Karl \cite{isg77,isg79} 
that the low--lying negative and positive parity non--strange baryon mass 
spectra can be 
well reproduced if the spin--orbit force from the OGE interaction is somehow
suppressed. In calculating the excited nucleon masses
Takeuchi found that the spin--orbit force from the 't~Hooft 
interaction mostly cancels the spin--orbit force originating in the OGE 
interaction. However, this cancellation does not occur in the 
nucleon--nucleon force where the combination of the OGE and 't~Hooft 
interactions produces a strong spin--orbit component. In the same work 
Takeuchi also examined the low--lying 
negative parity $N^*$ and $\Delta^*$ mass spectra using the MIT bag model 
with both OGE and 't~Hooft interactions and found that the cancellation 
seem to persist for bag models as well \cite{tak94}.
Unfortunately, however, both Blask {\em et al.} and Takeuchi focused 
only on the masses 
of the excited states and did not investigate the spin--flavor 
compositions of the baryon wave functions. In this work close attention 
is paid to the spin--flavor contents of excited baryons since they determine 
the meson--baryon 
vertex functions as well as radiative transition amplitudes to the ground 
states.

Kim and Banerjee developed a program to incorporate the 't~Hooft interaction 
into the CCM \cite{kim93} and their method is used in this work. They 
pointed out that although the strength of the 
't~Hooft interaction in QCD can be estimated using Shuryak's instanton 
liquid model of the QCD vacuum \cite{shu88}, the result cannot be used 
directly in the 
CCM where the block--spinning procedure of Nielsen and 
P\`{a}tkos~\cite{nie82} 
modifies the effective 't~Hooft interaction Lagrangian. Because of this reason 
the strength parameter $C_s$ was fixed by 
fitting the $N-\Delta$ mass splitting including both the `t Hooft 
and perturbative one--pion exchange (OPE) interactions. The results 
for the mass and the size of the nucleon showed remarkable improvement 
relative to earlier CCM results as mentioned in the Introduction.

However, the neutron charge radius turned out to be too small. 
Isgur {\it et al.}~\cite{ikk78_80,iks81} had pointed out that 
non--perturbative effects 
of OGE can explain the size of $\langle r^2\rangle_{\rm neutron}$. Similar 
effects exist in the CCM when the 't~Hooft interaction is used in 
conjunction with perturbative OPE interaction as the 
source of spin and isospin dependent forces.\footnote{Recall that in the 
CCM contributions from gluon exchanges, including the OGE, is effectively
included in the OPE interaction} The nucleon problem was redone in the CCM in 
\cite{kim94} by mixing all seven possible configurations made out 
of $1S_{\frac12}$, $1P_{\frac12}$ and $1P_{\frac32}$ single quark states. 
In addition, two configurations of the type $(1S_{\frac12})^2 2S_{\frac12}$ 
and one of the type $(1S_{\frac12})^2 1D_{\frac32}$ have 
also been used. The calculation yielded a very good fit for 
$\langle r^2\rangle_{\rm neutron}$.\footnote{Furthermore, with the pure mass 
$\chi$ potential used in this work, 
the coupling constant $g_{\pi NN^*}$ was largest for the lowest $N^*$ state 
and sharply smaller for the remaining ones.} 

For $N_f$ flavors the 't~Hooft interaction consists of a sum of one, 
two and up to  
$N_f$--body terms\cite{shi80}. For 3 flavors the interaction is given by
\begin{eqnarray}
{\cal L}_{\rm 't~Hooft} 
& = & 
-am_1m_2m_3\sum_i\bar{\psi}_{iR}\psi_{iL}/m_i+ 
b \biggl(m_3\bar{\psi}_{1R}\psi_{1L}\,\bar{\psi}_{2R}\psi_{2L} + 
{\rm permut.} \biggr)
\nonumber \\
&    & + c\bar{\psi}_{1R}\psi_{1L}\bar{\psi}_{2R}\psi_{2L}\bar{\psi}_{3R}
\psi_{3L} \nonumber \\
&    & + \Biggl[ \biggl(Am_3-A'\bar{\psi}_{3R}\psi_{3L}\biggr) 
\bar{\psi}_{1R}\lambda^a\psi_{1L}\,
\bar{\psi}_{2R}\lambda^a\psi_{2L} \Biggr]
\nonumber \\
&    & + Bd^{abc} \Biggl[\bar{\psi}_{1R}\sigma_{\mu\nu}\lambda^a
\psi_{1L}\bar{\psi}_{2R}
\sigma_{\mu\nu}\lambda^b\psi_{2L}\bar{\psi}_{3R}\lambda^c\psi_{3L} + {\rm 
permut.} \Biggr]
\nonumber \\
&    & + Cd^{abc}\bar{\psi}_{1R}\lambda^a\psi_{1L}\bar{\psi}_{2R}
\lambda^b\psi_{2L}
\bar{\psi}_{3R}\lambda^c\psi_{3L}
\nonumber \\
&    & + Df^{abc}\bar{\psi}_{1R}\sigma_{\mu\nu}\lambda^a\psi_{1L}
\bar{\psi}_{2R}
\sigma_{\nu\gamma}\lambda^b\psi_{2L}\bar{\psi}_{3R}\sigma_{\gamma\mu}
\lambda^c\psi_{3L}. 
\label{thooft3} 
\end{eqnarray} 
where $i=1, 2, 3$ refer to the three flavors and $\psi_{i R/L}$ is the
right/left--handed quark spinor with flavor $i$ and current quark mass 
$m_i$. The constants, $a$, $b$, $c$, $A$, $A'$, 
$B$, $C$ and $D$ depend upon the distribution of instanton size in the  
instanton liquid vacuum. 
When applied to the CCM the one--body term induces a change in the current 
quark mass of about 7\%. In practice, this small modification 
is ignored and Kim and Banerjee retained only the dominant two--body 
interaction for the 
description of the $N - \Delta$ system \cite{kim93}. 
The same apporach was used by Kochelev \cite{koc85} and 
Takeuchi \cite{tak94}, all of whom used 
bag models to calculate the effects of the 't~Hooft interaction 
on baryon masses. 

Using Fierz transformations the two--body 't~Hooft interaction in the CCM 
may be written in the SU(3) limit as \cite{kim93},
\begin{eqnarray}
H_{\rm 't~Hooft} 
& = &  \left( -\frac{2}{3} C_s \right) \sum_{i < j} \int\!d^3r \Bigg[
\overline{\psi}(i) \psi(i) \overline{\psi}(j) \psi(j ) + 
\overline{\psi}(i) \gamma_5 \psi(i) \overline{\psi}(j)  \gamma_5 \psi(j) 
\nonumber\\
&    & \;\;\;\;\;\;\;\; 
+ \overline{\psi}(i) \lambda^a(i) \psi(i) \overline{\psi}(j) 
\lambda^a(j)  \psi(j ) + 
\overline{\psi}(i) \gamma_5  \lambda^a(i) \psi(i) \overline{\psi}(j)  
\gamma_5  \lambda^a(j) \psi(j) 
\Biggr].
\label{eq:Hooft}
\end{eqnarray} 
As mentioned earlier the parameter determining the interaction strength 
$C_s$ is taken to be 1.6 ${\rm fm}^2$. 
In the mean field approximation the two--body attractive 't~Hooft interaction 
is ultraviolet 
divergent due to the fact that quark fields of different flavors are at 
the same space--time point, and leads to a collapsed baryon.
In this work the same regulator as in Kim and Banerjee \cite{kim93} is 
used to control this divergence with the same parameter of $r_c$ = 0.25 fm. 

Note that this two--body interaction 
acts only on flavor anti--symmetric quark pairs. Being a contact 
interaction the spatial state must be a $S$ state to receive 
any contribution. 
Then total spin--flavor--space 
symmetry requires that the quark pairs be in spin anti--symmetric state. In 
a flavor decuplet every quark pair is in spin symmetric state 
and the 't~Hooft 
interaction makes no contribution. However, in an octet quark pairs are 
in spin anti--symmetric state half the time and the 't~Hooft 
interaction contributes to 
its mass.\footnote{$H_{\rm 't~Hooft}$ also contributes to flavor singlet 
states as well.} This contribution is attractive, thus 
lowering the mass of the 
nucleon relative to that of the $\Delta$, helping to explain the 
observed $N - \Delta$ 
mass splitting. It should be noted that the pattern of the 
mass splitting induced by the 't~Hooft interaction is different from the 
one obtained by using the traditional OGE interaction. The reason is that 
while the 't~Hooft interaction does not contribute when the quark pair is 
in spin symmetric state, OGE produces a repulsive force 1/9 in magnitude 
of the attractive force in the spin anti--symmetric state. 

For $N_f = 3$, a three--body interaction among the differently 
flavored quarks are also possible in addition to the one-- and two--body 
interactions. 
However, for baryons such a three--body force is absent for the 
following reason \cite{oka89,bla90}. The 't~Hooft interaction effectively 
describes
the {\em change} of the eigenvalue of the zero mode due to the interaction 
between the quarks and the instanton liquid vacuum. In order to put three 
quarks in the initial zero mode they must be in a totally 
anti--symmetric state
in flavor and for these quarks to feel the three--body contact interaction 
the spatial wave function of the baryon must be totally symmetric. Thus the
spin wave function for the baryon must be totally anti--symmetric according
to total spin--flavor--space symmetry. However this is not possible with 
spin 1/2 quarks and the three--body 't~Hooft interaction  does not 
contribute for 
hyperons.\footnote{Nevertheless, for two hyperon systems the three--body
interaction does not vanish and it has been speculated that such an 
interaction may even 
overcome the attractive OGE interaction to unbind the hypothesized $H$ 
di--baryon \cite{oka89}.} 

\section{DIAGONALIZATION OF $H_{\rm Bare}$}
\subsection{Projection of the Translational Modes}
\label{diagonal}

The Hamiltonian used to evaluate the bare baryon masses and 
three--quark wave functions in this work is given by,
\begin{equation}
H_{\rm Bare} = H^{(0)}_{\rm Toy} + H_{\rm 't~Hooft}.
\label{eq:BARE}
\end{equation}
The matrix elements of $H_{\rm Bare}$ are evaluated in the $j-j$ 
coupled basis 
constructed by coupling the total spin $j = l + s$ of
individual quarks.\footnote{Lower case letters for total spin $j$, orbital 
angular momentum $l$ and intrinsic spin $s$ 
are used to label quarks while upper case letters are used for baryons.}
For negative parity baryons they are labeled as $|SU(3),J;SSX\rangle$ where 
$S$ is a quark in the ground $S_{1/2}$ state and $X$ can be either in the 
$P \equiv P_{1/2}$ or the $A \equiv P_{3/2}$ state. 
Before proceeding to the evaluation of the matrix elements of 
$H_{\rm Bare}$, it is necessary to identify and project out as much 
as possible the spurious components in the 
excited baryon wave functions corresponding to center--of--mass excitations. 
The prescription used in this work to do this has been used previously 
\cite{deg76b,myh84,umi89} and is strictly valid only in the non--relativistic
and SU(6) limits.

In the non--relativistic limit the spin--flavor wave functions
are constructed in the $L-S$ basis by coupling the relative orbital 
angular momentum 
of the three--quark system, $L$, to its total intrinsic spin $S$. 
The $L-S$ coupled basis, $|SU(6),^{2S+1}SU(3)_J\rangle$, are labeled by 
the SU(6) spin--flavor multiplet, the SU(3) flavor multiplet, total 
angular momentum $J = L + S$ and the spin multiplet $2S+1$ of 
the three--quark system. The resulting spin--flavor wave functions of 
low--lying negative parity baryons 
belong either to the 56 or the 70 representation of SU(6) spin--flavor. 
Totally symmetric 56--plet wave functions share the same 
permutational symmetry as the ground state spin--flavor wave functions. 
In the non--relativistic quark model with harmonic oscillator confining 
potential the negative parity 56--plet 
correspond to the translational mode of the three--quark system. 
The genuine excitations 
are represented by permutationally mixed symmetric 70--plet wave functions. 

Motivated by this example in the non--relativistic limit, 
the projection prescription 
assumes that those linear combinations in the $j-j$ basis corresponding to
the 56--plet $L-S$ coupled wave functions are the translational 
modes of the baryon and is thus spurious. Therefore, $H_{\rm Bare}$ 
is diagonalized 
by using only those linear combinations of the $j-j$ basis corresponding 
to the 
70--plet in the $L-S$ basis. Consequently, the spin--flavor compositions of 
negative parity baryons are described by using the $L-S$ coupled basis of 
the form 
$|70,^{2S+1}SU(3)_J\rangle$ as shown in the fourth column of Table~III. 
Also shown in the table are the spin--parity and masses of well--established
low--lying negative parity baryons examined in this and the following paper.

Appendix~A 
presents the result of $L-S$ to $j-j$ basis recoupling which is used 
to express the 
56 and 70 components of the $L-S$ basis in terms of the $j-j$ basis. 
Supurious states are mixed into hyperon wave functions when SU(6) symmetry is
broken by $m_s - m_{u,d}$ mass difference. 
The relative amount of spurious component contained in hyperon wave 
functions for 
different values of strange quark masses is estimated in Appendix~B. 
It is found that the spurious components in hyperon wave functions 
increases with 
the strange quark mass indicating that this prescription
is inapplicable for descriptions of baryons containing a heavy 
quark \cite{umi89}.
\subsection{Matrix Elements of $H_{\rm Bare}$}
\label{matrix}

Before presenting the results for bare masses and three--quark wave 
functions it is 
instructive to study the matrix elements of $H^{(0)}_{\rm Toy}$ and 
$H_{\rm 't~Hooft}$. Since the calculation itself is straightforward, 
but quite lengthy 
and not very informative, only the results are discussed in detail. 
The matrix 
elements of the Hamiltonian in the $L-S$ coupled basis are labeled as
\begin{equation}
H \equiv \left( 
\begin{array}{ccc}
\langle 70,^2Y_J|H| 70,^2Y_J \rangle & \langle 70,^48_J |H| 70,^2Y_J\rangle 
& \langle 70,^28_J |H| 70,^2Y_J \rangle \\
\langle 70,^48_J |H| 70,^2Y_J \rangle & \langle 70,^48_J |H| 70,^48_J \rangle 
& \langle 70,^48_J |H| 70,^28_J\rangle\\
\langle 70,^28_J |H| 70,^2Y_J \rangle & \langle 70,^28_J |H| 70,^48_J \rangle 
& \langle 70,^28_J |H| 70,^28_J \rangle
\end{array}
\right).
\label{eq:MATRIX}
\end{equation}
Henceforth the SU(6) spin--flavor label in the $L-S$ basis is dropped and 
will always be understood to be the 70--plet. $Y$ can be either 
a flavor singlet 
for the $\Lambda^*$ or a decuplet for the $\Sigma^*$ states. 
The mass matrices for $J^P = 3/2^-$ and $1/2^-$ 
$N^*$ states are given by the lower right $2 \times 2$ submatrix of 
Eq.~(\ref{eq:MATRIX}) 
involving only the flavor octet matrix elements. 

Table~\ref{Melements} shows the matrix elements of $H^{(0)}_{\rm Toy}$ and 
$H_{\rm 't~Hooft}$ for negative parity nucleon and hyperon resonances.
Because the 't~Hooft interaction does not contribute to the $\Delta^*$ 
states their bare masses are simply given by the sum of single quark 
energies plus the constant contribution from the classical $\chi$ field 
energy.\footnote{The same applies to the flavor decuplet components
of $\Sigma^*$ states.} Analytical expressions 
for the matrix elements of $H_{\rm 't~Hooft}$ are given in Appendix~C. 
The fourth column of the table shows the 
corresponding matrix elements of the OGE interaction evaluated in the
MIT bag model with vanishing quark masses for comparison. 

Note that in the SU(6) limit simple patterns emerge among the matrix elements:
\begin{itemize}
\item[] In the hyperon sector the model Hamiltonian does not connect 
flavor singlet or decuplet states with flavor octet states. As a result 
there are two pure flavor singlet $\Lambda^*$ states with $J^P=3/2^-$ 
and $1/2^-$. Correspondingly, there are two $\Sigma^*$ states 
which are purely flavor decuplets with $J^P=3/2^-$ and $1/2^-$. 
Since the 't~Hooft interaction does not 
act on flavor decuplets, the bare masses of these two $\Sigma^*$ states 
are given by the diagonal matrix elements of $H^{(0)}_{\rm Toy}$ 
of 1497 MeV and 1477 MeV, respectively.
\item[] These values of decuplet $\Sigma^*$ masses also correspond to that of 
$J^P = 3/2^-$ and $1/2^-$
$\Delta^*$ baryons resulting in two sets degenerate $\Sigma^*$ and 
$\Delta^*$ states. However, there are no degenerate partners in the 
non--strange 
sector for the purely flavor singlet $\Lambda^*$ states since no 
permutationally 
anti--symmetric three--quark flavor wave function can be constructed 
with two flavors.
\item[] The matrix elements of $H^{(0)}_{\rm Toy}$ and 
$H_{\rm 't~Hooft}$ involving flavor octet states 
$|^48\rangle_J$ and $|^28\rangle_J$ are the same for the $N^*$, 
$\Lambda^*$ and $\Sigma^*$ baryons sharing the same spin and parity. 
Thus the bare CCM 
Hamiltonian produces two sets of degenerate $N^*$, $\Lambda^*$ and 
$\Sigma^*$ states 
with $J^P = 3/2^-$ and another two degenerate sets with $J^P = 1/2^-$ in the 
SU(6) limit. 
\end{itemize}
These mass degeneracies are lifted when single particle  
energies of strange quarks are corrected for flavor symeetry breaking as 
discussed in Section~\ref{toymodel}. However since this correction is applied 
perturbatively it does not affect the hyperon wave functions.

Note that the off--diagonal matrix elements of $H_{\rm 't~Hooft}$ for 
$J^P=3/2^-$ states are neglegibly small while those for $J^P=1/2^-$ states 
are quite large inducing large mass splittings between the flavor octet 
states. When compared to the matrix elements of OGE interaction in the
MIT bag model one apparent difference
can be seen in the decuplet matrix elements involving the two $\Sigma^*$ 
states.  
Morever, for the flavor singlet 
matrix elements the 't~Hooft interaction gives a much larger and 
negative value 
of $-90$ MeV for the $J^P=3/2^-$ state compared with $-10$ MeV for the 
$J^P=1/2^-$ 
matrix element. This pattern is reversed for the OGE interaction where 
the singlet matrix element of the $J^P=1/2^-$ state of $\alpha_s (-61)$ MeV
is more negative than the $J^P=3/2^-$ matrix element of $\alpha_s (-12)$ MeV.

\subsection{Bare Masses and Three--Quark Wave Functions}
\label{baremass}
Bare masses and three--quark wave functions of negative parity baryons  
$N^*$, $\Delta^*$, $\Lambda^*$ and $\Sigma^*$ in the CCM are given in 
Table~\ref{Results} which lists the results before and after including 
the two--body 
't~Hooft interaction in the Hamiltonian. With the present set of 
model parameters the 't~Hooft interaction affects the masses of 
baryon resonances 
more than their spin--flavor compositions, the latter being determined mostly 
by single particle energies of valence quarks. The fourth column also shows 
the bare masses of hyperons after being corrected for flavor 
symmetry breaking. 

\subsubsection{$J^P=5/2^-$ States}

The matrix elements of $H^{(0)}_{\rm Toy}$ for $J^P=5/2^-$ 
$N^*$, $\Lambda^*$ and $\Sigma^*$ states are all equal and 
has the value of 1470 MeV. These baryons are purely flavor octet states. 
In the 
present quark shell model the 
three--quark wave functions for $J^P = 5/2^-$ baryons are constructed 
by exciting 
one of the ground state quarks from the $S$ to the excited $A$ state. 
Thus the contribution from single quark energies to $H^{(0)}_{\rm Toy}$ is
simply $2\epsilon_S + \epsilon_A$ = 1170 MeV while the
classical $\chi^{(0)}$ field energy contributes the remaining 300 MeV.
There is no flavor--decuplet $J^P=5/2^- \Delta^*$ state 
which is interpreted as a spurious excitation according to the 
center--of--mass 
projection prescription as shown in Eq.~(\ref{eq:JJLSa}). 
The absence of this $\Delta^*$ state is in agreement with experiment 
(see Table~III).

Note that numerically, the matrix elements of $H^{(0)}_{\rm Toy}$ and 
$H_{\rm Bare}$ are practically equal implying that the effects of 't~Hooft 
interaction on $J^P=5/2^-$ states are neglegiblely small. 
This is because the spin--isospin dependent 't~Hooft interaction 
acts on quark pairs where one of them is in the $S$ and the other in the 
$A$ state. 
Being a contact interaction, the resulting interaction energy is 
expected to be small as verifed numerically in this model calculation. 
Corrections from flavor symmetry breaking lifts the mass degeneracy 
between the well 
established $\Lambda(5/2^-)$ and $\Sigma(5/2^-)$ states \cite{isg78}. 
However, the resulting mass splitting is only 14 MeV in contrast to the 
empirical values of 55 MeV. Thus, a further mass 
splitting of about 40 MeV is necessary from meson exchange corrections.

\subsubsection{$J^P=3/2^-$ States}

For $J^P=3/2^-$ baryons there are three sets of degenerate eigenstates of 
$H_{\rm Bare}$. 
\begin{itemize} 
\item[ ] Purely flavor decuplet $\Sigma(3/2^-)_1$ and $\Delta(3/2^-)$ with 
$E_{\rm Bare}$ = 1497 MeV.
\item[ ] $N(3/2^-)_1, \Lambda(3/2^-)_1$ and $\Sigma(3/2^-)_2$ with 
$E_{\rm Bare}$ = 1485 MeV. The spin--flavor composition of these
states is approximately 72 \% $|^48_{3/2} \rangle$ and 28 \% $|^48_{3/2} 
\rangle$.
Using the results of $L-S$ to $j-j$ recoupling given in Appendix~A, 
this corresponds to 8\% $|8,3/2;SSA\rangle$, 14\% $|8,3/2;SSA\rangle'$ and 
78\% $|8,3/2;SSP\rangle$ in terms of $j-j$ coupled basis. 
Thus the excited valence quark in these baryons is mostly in the $P$ orbit 
with about 20\% probability of being in the $A$ orbit. 
\item[ ] $N(3/2^-)_2, \Lambda(3/2^-)_2$ and $\Sigma(3/2^-)_3$ with 
$E_{\rm Bare}$ = 1439 MeV. In this set of degenerate states, the excited 
quark is almost always in the $A$ state with a vanishingly small 
$|8,3/2;SSP\rangle$ 
component. Recall that the single quark energy for the $A$ state
($\epsilon_A$ = 490 MeV) is smaller than that of the $P$ state 
($\epsilon_P$ = 550 MeV).
This explains qualitatively why this set of degenerate states has a lower 
mass than the above set where the excited quark is dominantly in the $P$ 
orbit.
\end{itemize}
Finally, the purely flavor singlet $\Lambda(3/2^-)_3$ with 
$E_{\rm Bare}$ = 1380 MeV has no mass degenerate partners. The values for 
$E_{\rm Bare}$ for the octet and singlet states are all lower than the values 
obtained by using $H_{\rm Toy}$ alone indicating that the 't~Hooft 
interaction is 
attractive for the $J^P=3/2^-$ baryons.  
\subsubsection{$J^P=1/2^-$ States}

As in the $J^P=3/2^-$ sector there are three sets of degenerate eigenstates 
of $H_{\rm Bare}$ for $J^P=1/2^-$ baryons.
\begin{itemize}
\item[ ] Purely flavor decuplet $\Sigma(1/2^-)_2$ and $\Delta(1/2^-)$ with 
$E_{\rm Bare}$ = 1477 MeV.
\item[ ] $N(1/2^-)_1, \Lambda(1/2^-)_1$ and $\Sigma(1/2^-)_1$ with 
$E_{\rm Toy}$ = 1529 MeV and $E_{\rm Bare}$ = 1742 MeV. The spin--flavor 
composition of these
states is about 86\% $|^48_{1/2}\rangle$ and 14\% $|^28_{1/2}\rangle$ which 
translates into 92\% $|8,1/2;SSP\rangle$ and 8\% combination of 
$|8,1/2;SSA\rangle$ 
and $|8,1/2;SSA\rangle'$ states. Thus, the excited quark is dominantly in 
the $P$ orbit.
\item[ ] $N(1/2^-)_2, \Lambda(1/2^-)_3$ and $\Sigma(1/2^-)_3$ with 
$E_{\rm Toy}$ = 1496 MeV and $E_{\rm Bare}$ = 1450 MeV. 
In this set the excited quark is about 45\% 
in the $|8,1/2;SSP\rangle$ state and 55\% in the $A$ state leading to a 
lower value of energy eigenvalue than the above degenerate set.
\end{itemize}
Note that for the latter two sets $E_{\rm Bare}$ = 1742 and 1450 MeV, 
respectively, 
implying that the 't~Hooft interaction induces a large mass splitting among 
the $J^P = 1/2^-$ baryons. This 
result can be anticipated by observing that the diagonal matrix elements 
of $H_{\rm 't~Hooft}$ for the octet $J^P=1/2^-$ states differ by an order 
of magnitude as well as by sign (178 and -12 MeV) as shown in Table~IV.
Finally, the pure flavor--singlet $\Lambda(1/2^-)_2$ state with 
$E_{\rm Bare}$ = 1520 MeV has no degenerate partners as in the 
$J^P = 3/2^-$ case.

\subsection{Model comparisons}
\label{comparisons}
It is interesting to compare the CCM results for the three--quark 
wave functions with the results obtained in the
MIT bag model using the OGE interaction. 
Table~\ref{Percent} presents the masses and relative percentages of 
spin--flavor contents of low--lying $J^P=3/2^-$ and $1/2^-$ 
$\Lambda^*$ states in the CCM and in the MIT bag model using respectively the 
't~Hooft and OGE matrix elements of Table~IV.\footnote{The 
MIT bag model results were obtained with $B^{1/4}$ = 145 MeV, $Z_0$ = 0.25, 
$\alpha_s$ = 1.5 and $m_u = m_d = m_s$ = 0 MeV.} The same center--of--mass
projection prescription is used for both calculations. The Table 
also presents 
the results obtained in the NRQM \cite{isg77} 
which includes explicit flavor symmetry breaking. 
These negative parity $\Lambda^*$ states are chosen since radiative decay 
widths of the lightest $J^P=3/2^-$ $\Lambda(1520)$ and $1/2^-$ 
$\Lambda(1405)$ hyperons are examined in the following paper. 

The results for the wave functions of $\Lambda(3/2^-)$ states are 
qualitatively 
similar for both the CCM and the MIT bag model, with the lightest $J^P=3/2^-$ 
hyperon being a flavor singlet state. Both models predict similar 
spin--flavor 
compositions for $\Lambda(3/2^-)_1$ and $\Lambda(3/2^-)_2$. 
A similar result is also given in the NRQM with the lightest hyperon state 
having 
a 81\% flavor--singlet component.  
It should be noted that in the bound state approach to the 
Skyrme model no bound state solution corresponding to this state exists 
with physically acceptable values of model parameters \cite{riska}.
Thus these models suggest that the well-established $\Lambda(1520)$ is a
genuine three--quark excited state with a dominant spin--doublet 
flavor--singlet component.

Differences in model predictions for hyperon wave functions can clearly be 
seen in the $J^P=1/2^-$ sector where the only agreement among the 
three models 
is the heaviest $\Lambda(1/2^-)_1$, which is mostly a $|^48_{1/2}\rangle$ 
state. 
In the CCM $\Lambda(1/2^-)_2$ is a pure flavor singlet state while it is a 
spin--doublet flavor--octet state in the MIT bag model. The NRQM finds 
$\Lambda(1/2^-)_2$ 
to be a mixture of 56\% $|^2 8_{1/2}\rangle$, 34\% $|^4 8_{1/2}\rangle$ 
and 10\% $|^2 1_{1/2}\rangle$. The lightest $J^P=1/2^-$ hyperon 
is found to be dominantly a $|^28_{1/2}\rangle$ state in the CCM, while it is 
a pure flavor singlet state in the MIT bag model in the limit of vanishing 
quark masses. However, when flavor symmetry is broken in the MIT bag model 
a non--negligible octet component is mixed into the $\Lambda(1/2^-)_3$ wave 
function \cite{umi93,deg76b}. In the NRQM this
hyperon resonance is predominantly a flavor singlet state.
In previous works on the $K^-p$ atom and $\bar{K}N$ scattering 
in the cloudy bag model \cite{vei85,zho88}, it has been assumed that the 
$\Lambda(1405)$ is a pure flavor singlet. The present example 
indicates that the cloudy bag model 
Hamiltonian must be properly diagonalized before neglecting the 
flavor octet component of the $\Lambda(1405)$ wave function.

\section{MESON--BARYON VERTEX FUNCTIONS}
\label{ffactors}

In this section three--quark wave functions obtained by diagonalizing
$H_{\rm Bare}$ are used to calculate meson--excited bare 
baryon vertex functions to lowest order in meson--quark coupling. 
Each of these vertex functions is a product of a coupling constant 
and a form factor when the baryons are on their mass--shell. 
Meson--bare baryon form factors for negative parity resonances are extracted 
by applying a method originally introduced for ground states 
\cite{wei84,bro75}. It will be shown that model 
predictions for the form factors have problems with their analytical 
structures and, consequently, 
only their values at zero three--momentum transfer are reliable. As examples,
axial coupling constants for excited nucleons are determined using the 
Goldberger--Trieman relation.

The equation of motion for the pseudoscalar meson fields $\phi^a(\vec{r}\:)$ 
corresponding to the SU(3) CCM Lagrangian given in Eq.~(\ref{eq:CCM}) is,
\begin{equation}
\left( \Box + m_{\phi}^2 \right) \phi^a (\vec{r}\:) = 
\frac{g_\phi}{(g_\chi \chi^{(0)}(r))^2} \bar{\psi}(\vec{r}\:) 
\biggl( i\lambda^a \gamma_5 
\biggr) \psi(\vec{r}\:) \equiv J^a_5(\vec{r}\:)_q.
\label{eq:MESON}
\end{equation}
In the present work, where the excited quark is either in the $P$ 
or the $A$ state, the meson--quark source current is given by
\begin{eqnarray}
J^a_5(\vec{r}\:)_q 
& = & \frac{g_\phi}{(g_\chi \chi^{(0)}(r))^2} \Bigg[
\bar{u}_S(\vec{r}\:) \biggl( i\lambda^a \gamma_5 \biggr) u_S(\vec{r}\:)
+ \bar{u}_P(\vec{r}\:) \biggl( i\lambda^a \gamma_5 \biggr) u_P(\vec{r}\:) 
\nonumber \\ 
&  & \;\;\;\;\;\;\;\;\;\;\;\;\;\;\;\;\;\;\;\;\;\;\;\;\;\;\;\;\;\;
+ \bar{u}_A(\vec{r}\:) \biggl( i\lambda^a \gamma_5 \biggr) u_A(\vec{r}\:)
+ \bar{u}_P(\vec{r}\:) \biggl( i\lambda^a \gamma_5 \biggr) u_A(\vec{r}\:)  
\Bigg]
\label{eq:QMSOURCE}
\end{eqnarray}
where the quark spinors for states $S$, $P$ and $A$ are given in 
Eqs.~(\ref{eq:SWAVE}) to (\ref{eq:AWAVE}), respectively. 

To lowest order in meson--quark coupling, the meson--excited bare 
baryon vertex function is given by the matrix element of the Fourier 
transform of Eq.~(\ref{eq:QMSOURCE}) in the static limit, 
\begin{equation}
V(BB'\phi)(|\vec{q}\:|) \equiv
\sum_{j=1}^{3}\langle (qqq)_{B'}| \int d^3r e^{+i\vec{q}\cdot\vec{r}} 
J^a_5(\vec{r}\:)_q^j | (qqq)_B \rangle,
\label{eq:FFACTOR}
\end{equation}
where $| (qqq)_B \rangle$ is a three--quark state corresponding to baryon $B$.
On the right side of Eq.~(\ref{eq:FFACTOR}), summation over 
the quark index $j$ is shown explicitly for the single quark operator 
$J^a_5(\vec{r}\:)_q^j$. In general, the initial and final baryons, $B$ and 
$B'$, 
may belong to different spin--parity states. However, in this work only those 
vertex functions involving initial and final states with the same $J^P$ value 
are considered. This means that when using $V(BB'\phi)$ to calculate 
one--meson loop contributions to baryon observables, the intermediate 
baryon state is restricted to share the same spin--parity as the initial 
and final baryons. 

The Fourier transformed meson--quark source current operator may be 
written as,
\begin{eqnarray}
\int d^3r e^{+i\vec{q}\cdot\vec{r}} J^a_5(\vec{r}\:)^j_q 
&  =  & 
F_{SS}^{l=1}(|\vec{q}\:|) {\cal O}_{SS}^{a, l=1} (\hat{q}\:)^j
+ F_{PP}^{l=1}(|\vec{q}\:|) {\cal O}_{PP}^{a, l=1} (\hat{q}\:)^j 
+ F_{AA}^{l=1}(|\vec{q}\:|) {\cal O}_{AA}^{a, l=1} (\hat{q}\:)^j 
\nonumber \\
&    & 
\;\;\;\;\;\;\;\;\;\;\;\;\;\;\;\;\;\;\;\;\;\;\;\;\;\;\;\;\;\;
+ F_{PA}^{l=1}(|\vec{q}\:|) {\cal O}_{PA}^{a, l=1} (\hat{q}\:)^j
+ F_{AA}^{l=3}(|\vec{q}\:|) {\cal O}_{AA}^{a, l=3} (\hat{q}\:)^j.
\label{eq:VFUNCTION}
\end{eqnarray}
Here ${\cal O}^{a,l}_{MN}(\hat{q}\:)^j$ is the $l$--th multipole single 
quark operator acting on the spin--flavor space of the $j$--th quark.
The subscripts $MN$ identifies various contributions from the meson--quark 
source current, Eq~(\ref{eq:QMSOURCE}), in an obvious manner.
These operators and their corresponding functions $F^l_{MN}(|\vec{q}\:|)$ are 
defined in Appendix~D. 
If the initial and final baryon states are restricted to share the same
$J^P$ value, the operators shown in Eq.~(\ref{eq:VFUNCTION}) are the only 
posssible multipole contributions in the 
present model with a quark shell model space consisting only 
of $S$, $P$ and $A$ states. 

Meson--baryon form factors are 
determined by defining a meson--baryon interaction at the {\em hadronic} level
and equating the Fourier transformed meson--quark and meson--baryon source 
currents\cite{wei84,bro75}. 
For spin 1/2 baryons assume a meson--baryon interaction 
of the form 
\begin{equation}
{\cal L}_{\phi B'B} = g_{\phi B'B} \bar{\Psi}(\vec{r}\:)_{B'} 
\bigl( i\phi^a\lambda^a\gamma_5 \bigr) \Psi(\vec{r}\:)_B,
\label{eq:MBARYON}
\end{equation}
where $\Psi(\vec{r}\:)_B$ is a single baryon field with total spin $J$ and 
parity 
$P$. The corresponding meson--baryon source current $J^a_5(\vec{r}\:)_{B}$ is 
\begin{equation}
J^a_5(\vec{r}\:)_{B} \equiv g_{\phi B'B} \bar{\Psi}(\vec{r}\:)_{B'} 
\biggl( i\lambda^a \gamma_5 \biggr) \Psi(\vec{r}\:)_{B}.
\label{eq:BMSOURCE}
\end{equation}

For example, the $BB'\phi$ form factor $G_{BB'\phi}(k)$ for ground 
state octet baryons $B$ and $B'$ with $|J^P,J_z\rangle = |\frac{1}{2}^+,
+\frac{1}{2}\rangle$ is determined as follows:
\begin{eqnarray}
\lefteqn{ \langle B'(1/2^+,+1/2) | 
\int d^3r e^{+i\vec{q}\cdot\vec{r}} J^a_5(\vec{r}\:)_{B} 
|B (1/2^+,+1/2) \rangle} \nonumber \hspace{3cm} \\
& = &
-i G_{BB'\phi}^{l=1}(|\vec{q}\:|) \frac{|\vec{q}\:|}{2 M_N}
\langle B'(1/2^+,+1/2) | \lambda^a \hat{q}\cdot\vec{\sigma} 
|B (1/2^+,+1/2) \rangle 
\nonumber \\
& = &
-i G_{BB'\phi}^{l=1}(|\vec{q}\:|) \frac{|\vec{q}\:|}{2 M_N}
\langle B'| \lambda^a  |B \rangle 
\nonumber \\
& = &
\sum_{j=1}^{3}\langle (qqq)_{B'}| \int d^3r e^{+i\vec{q}\cdot\vec{r}} 
J^a_5(\vec{r}\:)_q^j | (qqq)_B \rangle  
\nonumber \\
& = &
\sum_{j=1}^{3}\langle (qqq)_{B'}| F^{l=1}_{SS}(|\vec{q}\:|)
{\cal O}^{a,l=1}_{SS}(\hat{q})^j| (qqq)_B \rangle.
\label{eq:MBFUN}
\end{eqnarray}
The superscript in $G_{BB'\phi}^{l=1}$ indicates that it is a 
dipole form factor. 
Note that by definition $G_{BB'\phi}^{l=1}$
is divided by twice the nucleon mass $2 M_N$ regardless of the types of 
initial and final baryons $B$ and $B'$. This convention, which differs 
from the one given in \cite{bro75} for the $N\Delta\pi$ form factor, 
will be extended to extract excited baryon form factors. 

For excited baryons with $|J^P,J_z\rangle = |\frac{5}{2}^-,
+\frac{5}{2}\rangle, |\frac{3}{2}^-,+\frac{3}{2}\rangle$, 
and $|\frac{1}{2}^-,+\frac{1}{2}\rangle$, it is 
found that only dipole form factors are numerically 
significant.\footnote{Form factors involving baryons belonging to different 
flavor multiplets may be obtained by exploiting the assumed SU(3) symmetry 
\cite{wei84,bro75}.} They are given by
\begin{eqnarray}
\lefteqn{ \underline{J^P=5/2^-:} \hspace{0.25 cm}
-i G_{BB'\phi}^{l=1}(|\vec{q}\:|) \frac{|\vec{q}\:|}{2 M_N} 
\left(\frac{1}{17}\right)
\langle B' |\lambda^a| B \rangle =} \hspace{4cm} \nonumber \\
&  &
\sum_{j=1}^{3}\langle (qqq)_{B'}| 
F^{l=1}_{SS}(|\vec{q}\:|){\cal O}^{a,l=1}_{SS}(\hat{q})^j
+ F^{l=1}_{AA}(|\vec{q}\:|){\cal O}^{a,l=1}_{AA}(\hat{q})^j
| (qqq)_B \rangle 
\label{eq:DFF52} \\
\lefteqn{ \underline{J^P=3/2^-:} \hspace{0.25 cm}
-i G_{BB'\phi}^{l=1}(|\vec{q}\:|) \frac{|\vec{q}\:|}{2 M_N} 
\left(\frac{1}{5}\right)
\langle B' |\lambda^a| B \rangle =} \hspace{4cm} \nonumber \\
&  &
\sum_{j=1}^{3}\langle (qqq)_{B'}| 
F^{l=1}_{SS}(|\vec{q}\:|){\cal O}^{a,l=1}_{SS}(\hat{q})^j
+ F^{l=1}_{PP}(|\vec{q}\:|){\cal O}^{a,l=1}_{PP}(\hat{q})^j \nonumber \\
&  & \;\;\;\;\;\;\;\;\;\;
+ F^{l=1}_{AA}(|\vec{q}\:|){\cal O}^{a,l=1}_{AA}(\hat{q})^j
+ F^{l=1}_{AP}(|\vec{q}\:|){\cal O}^{a,l=1}_{AP}(\hat{q})^j
| (qqq)_B \rangle 
\label{eq:DFF32}\\
\lefteqn{\underline{J^P=1/2^-:} \hspace{0.25 cm}
-i G_{BB'\phi}^{l=1}(|\vec{q}\:|) \frac{|\vec{q}\:|}{2 M_N} 
\langle B' |\lambda^a| B \rangle =} \hspace{4cm} \nonumber \\
& &
\sum_{j=1}^{3}\langle (qqq)_{B'}| 
F^{l=1}_{SS}(|\vec{q}\:|){\cal O}^{a,l=1}_{SS}(\hat{q})^j
+ F^{l=1}_{PP}(|\vec{q}\:|){\cal O}^{a,l=1}_{PP}(\hat{q})^j \nonumber \\
&  & \;\;\;\;\;\;\;\;\;\;
+ F^{l=1}_{AA}(|\vec{q}\:|){\cal O}^{a,l=1}_{AA}(\hat{q})^j
+ F^{l=1}_{AP}(|\vec{q}\:|){\cal O}^{a,l=1}_{AP}(\hat{q})^j
| (qqq)_B \rangle 
\label{eq:DFF12}
\end{eqnarray}

Figure 2 shows the normalized form factors $R(N^*N^*\pi)(|\vec{q}\:|) 
\equiv G^{l=1}_{N^*N^*\pi}(|\vec{q}\:|)/G^{l=1}_{N^*N^*\pi}(0)$
for negative parity nucleons under consideration. It can be seen from
the figure that all the $N^*N^*\pi$ form
factors are qualitatively similar to lowest order in meson--quark coupling.
In particular, the three momentum cut--off of each of the form factors
is around $|\vec{q}\:|$ = 1200 MeV. In the hyperon sector, a similar
$|\vec{q}\:|$ dependence can also be seen in the normalized dipole form
factors involving the $J^P=3/2^-$ flavor singlet and $N^*$ states,
$R(\Lambda_1 N^* k)(|\vec{q}\:|)$, as shown in Figure~3a. However,
the corresponding form factors involving the $J^P=1/2^-$ states, shown in
Figure~3b, are quite different. In this case the $\Lambda_1 N^* k$ form
factors change sign for the first time around $|\vec{q}\:| = 400$ MeV and
encounter a second node at about $|\vec{q}\:|$ = 1200 MeV. As the figure 
indicates these two form factors are proportional to each other 
since the Fourier trasformed meson--quark source current only connects 
the flavor singlet $\Lambda(1/2^-)_1$ to the spin--doublet flavor--octet 
($|^28_{1/2}\rangle$) component of the excited nucleon $N(1/2^-)$. 

Having extracted the meson--baryon form factors from meson--quark 
vertex functions 
it is tempting to calculate the meson--baryon coupling constants. However it
should be emphasized that the model form factors presented in this section 
do not have the correct behaviour when analytically continued 
to the complex $|\vec{q}\:|$--plane. The analytically continued CCM 
form factors 
are entire functions on the {\em whole complex plane} and not the cut plane 
with a branch point at the normal threshold of $|\vec{q}\:|^2 = 4M^2_B$ 
\cite{bar65}. 
In particular, there are no dynamical singularities in the CCM form factors 
which arise from summing over intermediate states in the dispersion 
representation. Thus the $|\vec{q}\:|$ dependencies 
of these form factors are questionable and the only reliable quatity to 
extract from these form factors are their values at $|\vec{q}\:|=0$, such as 
$G_{NN\pi}(0)$ to be discussed below. In order to 
evaluate the meson--baryon coupling constants the value of the form factor 
at $|\vec{q}\:|^2=m^2_\phi$ is required. For pion--baryon coupling constants 
it might be reasonable to Taylor expand the form factor around 
$|\vec{q}\:|=0$ since the pion mass is small,
\begin{equation}
G_{BB'\pi}(m_\pi^2) = G_{BB'\pi}(0) + m_\pi^2 
\frac{d}{d|\vec{q}\:|^2}G_{BB'\pi}(|\vec{q}\:|^2)|_{|\vec{q}\:|^2=0} 
+ \cdot\cdot\cdot.
\end{equation}
However this expansion would certainly not be justified for kaons and other 
heavier mesons. It should be remarked that
this problem is shared by other relativistic quark models with confinement
and meson--quark coupling, such as the chiral bag model \cite{wei84}.

When $B = B' = N(1/2^+)$ and $\phi = \pi$, the value of the CCM hadronic
form factor at $|\vec{q}\:|=0$ is given by 
$G_{N(1/2^+)N(1/2^+)\pi}^{l=1}(0) = 12.1.$ 
For excited nucleons the values of dipole form factors 
$G_{N^*N^*\pi}^{l=1}(|\vec{q}\:|)$ 
at zero three momentum transfers are found to be 
\begin{equation}
G_{N(5/2^-) N(5/2^-)\pi}^{l=1}(0) = 157.5
\end{equation}
\begin{equation}
G_{N(3/2^-)_1 N(3/2^-)_1\pi}^{l=1}(0) = 41.3 \hspace{1 cm}
G_{N(3/2^-)_2 N(3/2^-)_2\pi}^{l=1}(0) = -16.6 
\end{equation}
\begin{equation}
G_{N(1/2^-)_1 N(1/2^-)_1\pi}^{l=1}(0) = 8.4 \hspace{1 cm}
G_{N(1/2^-)_2 N(1/2^-)_2\pi}^{l=1}(0) = -2.4.
\end{equation}
The reason for large values of $G_{N(5/2^-) N(5/2^-)\pi}^{l=1}(0)$ and
$G_{N(3/2^-)_1 N(3/2^-)_1\pi}^{l=1}(0)$ are due to the spin matrix 
elements of 
(1/17) and (1/5) appearing on the left sides of Eqs.~(\ref{eq:DFF52}) 
and (\ref{eq:DFF32}), respectively. 

If the Goldberger--Trieman relation
\begin{equation}
g_A(0) = \frac{f_\pi}{M_N} G_{NN\pi}(0)
\label{eq:GTR}
\end{equation}
is applied to the ground state form factor $G_{N(1/2^+)N(1/2^+)\pi}^{l=1}$, 
the resulting axial coupling 
constant is $g_A \approx 1.21$ which is close to the empirical value of 
$g_A \approx 1.26$. Note that in the derivation of Eq.~(\ref{eq:GTR}), 
the free Dirac equation is used to introduce the nucleon mass $M_N$. 
When this relation is extended to 
excited states, the nucleon mass $M_N$ in Eq.~(\ref{eq:GTR}) should be 
replaced with the excited nucleon masses $M_N^*$, {\em i.e.}
\begin{equation}
g_A(0) = \frac{f_\pi}{M_N^*}G_{N^*N^*\pi}(0).
\end{equation}
Thus to lowest order in meson--quark coupling, the axial coupling constants 
for the excited nucleons are 
\begin{equation}
g_A^{N(5/2^-)} = 8.74
\end{equation}
\begin{equation}
g_A^{N(3/2^-)_1} = 2.3 \hspace{1 cm}  g_A^{N(3/2^-)_2} = -1.0
\end{equation}
\begin{equation}
g_A^{N(1/2^-)_1} = 0.47 \hspace{1 cm} g_A^{N(1/2^-)_2} = -0.15
\end{equation}

\section{DISCUSSION}
\label{discussion}

In this paper a toy model version of CCM is 
used together with the classical 't~Hooft interaction to calculate the quark 
contribution to the masses and wave functions of low--lying negative parity 
baryons $N^*, \Delta^*, \Lambda^*$ and $\Sigma^*$.
A baryon in the CCM is a composite object consisting of interacting quarks, 
mesons and a chiral singlet $\chi$ field confined inside a dynamically 
generated region. The meson fields in this model 
are effective degrees of freedom describing the exchange of a tower of 
gluons between quarks. 
In the mean field approximation which is used in this work, meson 
exchanges between quarks are treated perturbatively at the quantum level. 
Hence in the mean--field theory there is a ``bare" baryon with its 
wave function determined by the valence quark dynamics which couples to 
mesons yielding a physical baryon state. Mesonic exchange corrections to bare 
baryon properties will be examined in a following paper.

Masses and three--quark wave functions of bare excited baryon states 
are calculated by diagonalizing $H_{\rm Bare}$ defined in  
Eq.~(\ref{eq:BARE}) using $m_u = m_d = m_s$ = 7.5 MeV. A prescription, 
strictly valid only in the SU(6) and non--relativistic limits, 
is used to approximately eliminate the 
spurious degrees of freedom associated with the center--of--mass motion 
of the baryon. A simple test showed that this prescription 
becomes less reliable with increasing strange quark mass. It is found 
that the 't~Hooft interaction affects the bare masses of negative 
parity baryons 
more than their wave functions which are determined mostly by single 
quark energies. The CCM results for the bare masses indicate
that mesonic corrections to excited baryon masses must play an 
important role in order to reproduce the observed mass splittings. 
It should be remarked that corrections to bare masses and wave functions 
due to flavor symmetry breaking should be investigated. 

Using the bare baryon wave functions meson--excited baryon vertex 
functions have been calculated to lowest order in meson--quark coupling. 
Meson--baryon form factors for negative parity baryons were extracted 
from vertex functions by extending the conventional definition used for 
ground states. However, it is emphasized that because the model form 
factors do not have the correct behaviour when analytically 
continued to the complex plane, their momentum dependencies are 
questionable and only their values at zero momentum transfer can be trusted. 
By applying the Goldberger--Trieman relationship the quark contribution 
to weak axial charges for negative parity nucleons have been determined. 
In a following paper the vertex functions are used to regulated 
mesonic corrections 
to bare baryon properties in the one--loop and heavy baryon approximations. 
The second paper also examines the 
four radiative decays of the two lightest $\Lambda^*$ hyperons which 
will be measured in an upcoming Jefferson Lab experiment \cite{CLAS},
and will serve as an excellent testing ground for {\em all} models of baryon 
structure.

\acknowledgements
This work is supported by the US Department of Energy under grant No. 
DE--FG02--93ER--40762. 
\appendix
\section{$L-S$ to $j-j$ basis recoupling}
\label{appendixa}
As discussed in Section~\ref{diagonal}, the center--of--mass projection 
prescription used in this work assumes that the spurious components in 
negative parity baryon wave functions are described by those linear 
combinations in the $j-j$ basis corresponding to the 56--plet in the $L-S$ 
basis \cite{deg76b}. The 
remaining linear combinations corresponding to the 70--plet
are used to diagonalize the model Hamiltonian. This Appendix presents the 
results of $L-S$ to $j-j$ recoupling using the notations for $L-S$ 
and $j-j$ coupled basis defined in Section~\ref{diagonal}.
\begin{eqnarray}
\underline{J^P=5/2^-} &    & \nonumber \\
|56,^410_{5/2}\rangle       & = & |10,5/2;SSA\rangle \label{eq:JJLSa} \\
|70,^48_{5/2}\rangle         & = & |8,5/2;SSA\rangle \label{eq:JJLSb} \\
                                 &    & \nonumber\\
\underline{J^P=3/2^-} &    & \nonumber \\
|70,^21_{3/2}\rangle       & = & -|1,3/2;SSA\rangle' \\
\left( 
\begin{array}{c}
|56,^410_{3/2}\rangle\\
|70,^210_{3/2}\rangle
\end{array}
\right) & = &
\left(
\begin{array}{cc}
\frac{2}{3} & \frac{\sqrt{5}}{3} \\
\frac{\sqrt{5}}{3} & -\frac{2}{3}
\end{array}
\right) 
\left( 
\begin{array}{c}
|10,3/2;SSA\rangle\\
|10,3/2;SSP\rangle
\end{array}
\right)
\\
\left(
\begin{array}{c}
|56,^28_{3/2}\rangle\\
|70,^48_{3/2}\rangle \\
|70,^28_{3/2}\rangle
\end{array}
\right) & = &
\left(
\begin{array}{ccc}
\sqrt{\frac{5}{18}} & \frac{1}{\sqrt{2}} & -\frac{\sqrt{2}}{3} \\
\frac{2}{3}                 & 0                                 & 
\frac{\sqrt{5}}{3} \\
-\sqrt{\frac{5}{18}} & \frac{1}{\sqrt{2}} & \frac{\sqrt{2}}{3} 
\end{array}
\right) 
\left( 
\begin{array}{c}
|8,3/2;SSA\rangle \\
|8,3/2;SSA\rangle' \\
|8,3/2;SSP\rangle
\end{array}
\right)
\\
                                 &    & \nonumber\\
\underline{J^P=1/2^-} &    & \nonumber \\
|70,^21_{1/2}\rangle       & = & -|1,1/2;SSP\rangle' \\
\left( 
\begin{array}{c}
|56,^410_{1/2}\rangle\\
|70,^210_{1/2}\rangle
\end{array}
\right) & = &
\left(
\begin{array}{cc}
\frac{1}{3} & \frac{\sqrt{8}}{3} \\
\frac{\sqrt{8}}{3} & -\frac{1}{3}
\end{array}
\right) 
\left( 
\begin{array}{c}
|10,1/2;SSA\rangle\\
|10,1/2;SSP\rangle
\end{array}
\right)
\\
\left(
\begin{array}{c}
|56,^28_{1/2}\rangle\\
|70,^48_{1/2}\rangle \\
|70,^28_{1/2}\rangle
\end{array}
\right) & = &
\left(
\begin{array}{ccc}
\frac{2}{3} & \frac{1}{\sqrt{2}} & -\frac{\sqrt{2}}{6} \\
\frac{1}{3}                 & 0  & \frac{\sqrt{8}}{3} \\
-\frac{2}{3} & \frac{1}{\sqrt{2}} & \frac{\sqrt{2}}{6} 
\end{array}
\right) 
\left( 
\begin{array}{c}
|8,1/2;SSA\rangle \\
|8,1/2;SSP\rangle' \\
|8,1/2;SSP\rangle
\end{array}
\right)
\end{eqnarray}

\section{A test for the projection prescription}
\label{appendixb}

Because the projection prescription assumes SU(6) symmetry, the 
matrix elements of $H_{\rm Bare}$, Eq.~(\ref{eq:BARE}), 
are evaluated using $m_s = m_{u,d}$. Corrections from flavor 
symmetry breaking ($m_s \neq m_{u,d}$) to
neative parity hyperon masses are given to first order as discussed in
Section~\ref{toymodel}. However, the breaking
of flavor symmetry introduces spurious components in excited hyperon 
wave functions. This Appendix presents an estimate of the relative amount of 
spurious component introduced in excited hyperon wave functions  
for different values of strange quark mass using the well--established 
$J^P=5/2^-$ $\Sigma(1775)$ resonance as an example.

According to the projection prescription the wave function of this hyperon 
resonance, $|\Sigma(1775)\rangle \equiv |\Sigma^* \rangle_{\rm ex}$, 
is a pure 
flavor octet state identified with the 70--plet in the $L-S$ coupled basis 
as shown in Eq.~(\ref{eq:JJLSb}). The flavor decuplet state given in 
Eq.~(\ref{eq:JJLSa}) is assumed to be a spurious state, 
$|\Sigma^*\rangle_{\rm sp}$, 
corresponding to the translational mode of the baryon. In the $j-j$ 
coupled basis these two orthogonal states are given by 
\begin{eqnarray}
|\Sigma^* \rangle_{\rm ex} & = & |8,5/2;SSA\rangle = 
\sqrt{\frac{2}{3}} |\Sigma^*\rangle_1 - \sqrt{\frac{1}{3}} 
|\Sigma^*\rangle_2, 
\label{eq:SIGMA} \\
|\Sigma^*\rangle_{\rm sp} & = & |10,5/2;SSA\rangle = 
\sqrt{\frac{1}{3}} |\Sigma^*\rangle_1 + \sqrt{\frac{2}{3}} 
|\Sigma^*\rangle_2, 
\label{eq:SPURIOUS}
\end{eqnarray}
where,
\begin{eqnarray}
|\Sigma^*\rangle_1 & \equiv & \frac{1}{\sqrt{6}}
\bigl( 
u\!\uparrow\! d\!\uparrow\! s\:\alpha +\; 
d\!\uparrow\! u\!\uparrow\! s\:\alpha +\; 
u\!\uparrow\! s\:\alpha\: d\!\uparrow +\; 
d\!\uparrow\! s\:\alpha\: u\!\uparrow +\; 
s\:\alpha\: u\!\uparrow\! d\!\uparrow +\; 
s\:\alpha\: d\!\uparrow\! u\!\uparrow 
\bigr), 
\label{eq:SIGMAA}\\
|\Sigma^*\rangle_2 & \equiv & \frac{1}{\sqrt{12}}
\bigl( 
u\!\uparrow\! d\:\alpha\: s\!\uparrow +\; 
d\:\alpha\: u\!\uparrow\! s\!\uparrow +\; 
u\!\uparrow\! s\!\uparrow\! d\:\alpha +\; 
d\:\alpha\: s\!\uparrow\! u\!\uparrow +\; 
s\!\uparrow\! u\!\uparrow\! d\:\alpha +\; 
s\!\uparrow\! d\:\alpha\: u\!\uparrow \nonumber\\
                            &           & \;\;\; +\;
u\:\alpha\: d\!\uparrow\! s\!\uparrow +\; 
d\!\uparrow\! u\:\alpha\: s\!\uparrow +\; 
u\:\alpha\: s\!\uparrow\! d\!\uparrow +\; 
d\!\uparrow\! s\!\uparrow\! u\:\alpha +\; 
s\!\uparrow\! u\:\alpha\: d\!\uparrow +\; 
s\!\uparrow\! d\!\uparrow\! u\:\alpha 
\bigr).
\label{eq:SIGMAB}
\end{eqnarray}
The spin states are defined as $\uparrow \equiv |j=1/2,j_z=+1/2\rangle$ 
and $\alpha \equiv |j=3/2,j_z=+3/2\rangle$. Note that in Eq.~(\ref{eq:SIGMA})
the strange quark is in the excited $A$ state with a probability 
of two--thirds,
while this number is lowered to one--third for the
spurious $|\Sigma^*\rangle_{\rm sp}$ state shown in Eq.~(\ref{eq:SPURIOUS}). 
The SU(6) spin--flavor wave function for the ground state hyperon 
$J^P=1/2^+\; \Sigma(1192)$, denoted as $|\Sigma^0\rangle_{\rm gs}$, is 
\begin{equation}
|\Sigma^0\rangle_{\rm gs} = \sqrt{\frac{2}{3}} 
|\Sigma^0\rangle_1 - \sqrt{\frac{1}{3}} |\Sigma^0\rangle_2,
\label{eq:SIGMA0}  
\end{equation}
where,
\begin{eqnarray}
|\Sigma^0\rangle_1 & \equiv & \frac{1}{\sqrt{6}}
\bigl( 
u\!\uparrow\! d\!\uparrow\! s\!\downarrow +\; 
d\!\uparrow\! u\!\uparrow\! s\!\downarrow +\; 
u\!\uparrow\! s\!\downarrow\! d\!\uparrow +\; 
d\!\uparrow\! s\!\downarrow\! u\!\uparrow +\; 
s\!\downarrow\! u\!\uparrow\! d\!\uparrow +\; 
s\!\downarrow\! d\!\uparrow\! u\!\uparrow 
\bigr),
\label{eq:SIGMA0A} \\
|\Sigma^0\rangle_2 & \equiv & \frac{1}{\sqrt{12}}
\bigl( 
u\!\uparrow\! d\!\downarrow\! s\!\uparrow +\; 
d\!\downarrow\! u\!\uparrow\! s\!\uparrow +\; 
u\!\uparrow\! s\!\uparrow\! d\!\downarrow +\; 
d\!\downarrow\! s\!\uparrow\! u\!\uparrow +\; 
s\!\uparrow\! u\!\uparrow\! d\!\downarrow +\; 
s\!\uparrow\! d\!\downarrow\! u\!\uparrow \nonumber\\
                            &           & \;\;\; +\;
u\!\downarrow\! d\!\uparrow\! s\!\uparrow +\; 
d\!\uparrow\! u\!\downarrow\! s\!\uparrow +\; 
u\!\downarrow\! s\!\uparrow\! d\!\uparrow +\; 
d\!\uparrow\! s\!\uparrow\! u\!\downarrow +\; 
s\!\uparrow\! u\!\downarrow\! d\!\uparrow +\; 
s\!\uparrow\! d\!\uparrow\! u\!\downarrow 
\bigr),
\label{eq:SIGMA0B} 
\end{eqnarray}
with $\downarrow \equiv |j=1/2,j_z=-1/2\rangle$.

Let ${\cal O}$ be some 
operator exciting the internal degrees of freedom of the three--quark 
system so that ${\cal O}|\Sigma^0\rangle_{\rm gs}$ is proportional to 
$|\Sigma^* \rangle_{\rm ex}$. 
Then the following relations hold between the ground and excited state 
wave functions in the exact SU(6) limit 
\begin{eqnarray}
_{\rm ex}\langle\Sigma^*| {\cal O} |\Sigma^0 \rangle_{\rm gs} & \neq & 0,
\label{eq:TESTA} \\
_{\rm sp}\langle\Sigma^*| {\cal O} |\Sigma^0 \rangle_{\rm gs} & = & 0. 
\label{eq:TESTB} 
\end{eqnarray}
The deviation from the vanishing value of Eq.~(\ref{eq:TESTB}) is used to 
estimate the amount of spurious component introduced due to flavor symmetry 
breaking as follows.

The excitation operator ${\cal O}$ may be constructed by considering two 
single particle operators $S^{SS}_\mu$ and $L^{AS}_\mu$ defined as
\begin{eqnarray}
S^{SS}_\mu  & \equiv & \int d^3\!r\; \overline{u}_S(\vec{r}\:) \gamma_5 
\gamma_\mu u_S(\vec{r}\:)
\label{eq:SSS} \\
L^{AS}_\mu	& \equiv & \int d^3\!r\; \overline{u}_A(\vec{r}\:) 
r_\mu u_S(\vec{r}\:) 
\label{eq:LAS}
\end{eqnarray}
where $u_S$ and $u_A$ are the quark spinors for states $S$ and 
$A$ shown in Eqs.~(\ref{eq:SWAVE}) and (\ref{eq:AWAVE}), respectively. 
Let $S_+ \equiv (S^{SS}_1 + iS^{SS}_2)/2$ and 
$L_+ \equiv (L^{AS}_1 + iL^{AS}_2)/2$.
The operator $L_+ S_+$ acts on the
spin--flavor space of a quark in the following way,
\begin{equation}
L_+S_+ |u\!\uparrow \rangle = 
L_+S_+ |d\!\uparrow \rangle = 
L_+S_+ |s\!\uparrow \rangle = 0, 
\label{eq:LSA}
\end{equation}
and
\begin{equation}
L_+S_+ |u\!\downarrow \rangle = L_0S_0 |u\alpha \rangle; \;\;\;
L_+S_+ |d\!\downarrow \rangle = L_0S_0 |d\alpha \rangle; \;\;\;
L_+S_+ |s\!\downarrow \rangle = L_mS_m |s\alpha \rangle.
\label{eq:LSB}
\end{equation}
In Eq.~(\ref{eq:LSB}) $S_0$, $L_0$, $S_m$ and $L_m$ are defined as
\begin{eqnarray}
S_{0/m} & \equiv & 4\pi\int d\!r r^2\Bigl(G_0^2(r) + 
\frac{1}{3} F_0^2(r) \Bigl)_{0/m}, \\
L_{0/m} & \equiv & \frac{4\pi}{3\sqrt{2}} \int d\!r r^3\Bigl(-G_0(r)G_3(r) + 
F_0(r)F_3(r) \Bigr)_{0/m}.
\end{eqnarray}
The value of the up and down quark masses used to obtain the radial functions 
$G_i$ and $F_i$ for $S_0$ and $L_0$ is 7.5 MeV, while the strange quark mass  
has been left as a parameter for $S_m$ and $L_m$. 

Thus the operator $L_+S_+$ raises the total angular momentum $j$ of a
quark of given flavor from $\downarrow$ to $\alpha$ while annihilating
a single quark state with $j$ = $\uparrow$. Using $L_+S_+$ the excitation 
operator ${\cal O}$ is constructed as 
\begin{equation}
{\cal O} \equiv \sum_{k=1}^3 L_+(k) S_+(k),
\label{eq:OPERATOR}
\end{equation}
where $k$ is the quark index. When ${\cal O}$ acts on the ground state wave 
function $|\Sigma^0 \rangle_{\rm gs}$ it excites quarks with 
$j$ = $\downarrow$ to 
$\alpha$ thus creating either the $J^P=5/2^-$ $|\Sigma^*\rangle_{\rm ex}$ or 
$|\Sigma^*\rangle_{\rm sp}$ state where the excited quark must be in the $A$ 
state. 

Taking $m_s - m_{u,d}$ mass difference into account, the matrix elements 
corresponding to Eqs.~(\ref{eq:TESTA}) and (\ref{eq:TESTB}) are found to be
\begin{eqnarray}
_{\rm ex}\langle\Sigma^*|{\cal O}|\Sigma^0 \rangle_{\rm gs} & = & 
\frac{2}{3}L_mS_m + \frac{1}{3}L_0S_0,
\label{eq:TESTAA} \\
_{\rm sp}\langle\Sigma^*|{\cal O}|\Sigma^0 \rangle_{\rm gs} & = & 
\frac{\sqrt{2}}{3}\biggl(L_mS_m - L_0S_0\biggr). 
\label{eq:TESTBB} 
\end{eqnarray}
When $m_s = m_{u,d}$, $L_m \rightarrow L_0$ and $S_m \rightarrow S_0$.
Consequently, Eqs.~(\ref{eq:TESTA}) and (\ref{eq:TESTB}) hold exactly 
in the SU(6) limit and 
$|\Sigma^*\rangle_{\rm ex}$ 
is interpreted as the genuinly excited $|\Sigma(1775)\rangle$ state. 
However when $m_s \neq m_{u,d}$, $L_mS_m - L_0S_0 \neq 0$ and 
Eq.~(\ref{eq:TESTBB}) no longer vanishes indicating the presence of spurious 
components in the excited ${\cal O}|\Sigma^0\rangle_{\rm gs}$ state. 

The difference 
$\delta LS \equiv |L_mS_m - L_0S_0|$, which may be calculated 
in any relativistic quark
model, is used as an estimate of theoretical uncertainty 
arsing from $m_s - m_{u,d}$ mass difference. The values for $\delta LS$ 
are found to be 0.151 and 0.206 for $m_s$ = 200 and 300   
MeV, respectively, indicating that the projection prescription becomes 
less applicable with increasing strange quark mass. Similar conclusion is 
presented in \cite{umi89} where it is remarked that MIT 
bag model and its chirally extended versions are not suitable for 
describing a baryon with one heavy quark. It should 
be remarked that the projection prescription assumes that the angular 
momentum of a quark spinor is given by its upper component and this 
assumption contributes to the model uncertainty even in the SU(6) limit.  

\section{'t~Hooft interaction matrix elements}
\label{appendixc}
This Appendix presents the matrix elements of $H_{\rm 't~Hooft}$, 
Eq.~(\ref{eq:Hooft}), 
used in calculating the bare masses of low--lying negative parity baryons. 
The matrix elements have been evaluated assuming $m_s = m_{u,d}$ and 
are to be multiplied by an universal factor of $-\frac{16\pi}{3}C_S$ 
with $C_S = 1.6\; {\rm fm}^2$ 
\cite{kim93,kim94}. As discussed in Section~\ref{Hooft}, the 
't~Hooft interaction 
needs to be regularized. The regulator used in this work is
\begin{equation}
R(r) = 1 - e^{(r/r_c)^2},
\label{eq:REGULATE}
\end{equation}
which is the same as in Kim and Banerjee \cite{kim93} with the same 
parameter $r_c$ = 0.25 fm. The notation for the $j-j$ coupled basis 
are defined in Section~\ref{diagonal}. 
Recall that since the two--body 't~Hooft 
interaction operates on flavor anti--symmetric quark pairs, 
the $\Delta^*$ states and flavor decuplet components
of $\Sigma^*$ are unaffected by this interaction. 
Also because of assumed flavor symmetry, the
matrix elements involving the octet components of $N^*$, 
$\Lambda^*$ and $\Sigma^*$ are equal for given spin and parity. 

\vskip 1cm 
\noindent{$\underline{J^P=5/2^-:N^*/\Lambda^*/\Sigma^*}$}
\begin{eqnarray}
\langle 8,5/2;SSA|H|8, 5/2;SSA\rangle  & =  & \frac{3}{2}{\rm F1} 
\end{eqnarray}
\noindent{$\underline{J^P=3/2^-:N^*/\Lambda^*/\Sigma^*}$}
\begin{eqnarray}
'\langle 1,3/2;SSA|H|1,3/2;SSA\rangle' & = & {\rm F1} + 
{\rm F3} + {\rm F5} \\  
\langle 8,3/2;SSA|H|8,3/2;SSA\rangle& = & \frac{3}{5}\Bigl(
\frac{3}{4}{\rm F1} - \sqrt{3}{\rm F2} 
+ \frac{3}{4}{\rm F3}+ {\rm F4} \Bigr)\\  
'\langle 8,3/2;SSA|H|8,3/2;SSA\rangle' & = & \frac{1}{4}\Bigl({\rm F1} 
+ {\rm F3} + 4{\rm F5}\Bigr)\\  
\langle 8,3/2;SSP|H|8,3/2;SSP\rangle & = & \frac{3}{2}{\rm F6} \\
'\langle 8,3/2;SSA|H|8,3/2;SSA\rangle & = & \frac{3}{2\sqrt{5}}\Bigl(
-\frac{1}{2}{\rm F1} - 
\frac{1}{\sqrt{3}}{\rm F2} + \frac{1}{2}{\rm F3}\Bigr)\\  
\langle 8,3/2;SSP|H|8,3/2;SSA\rangle & = & \frac{3}{\sqrt{10}}\Bigl(
-\frac{\sqrt{3}}{2}{\rm F7} + {\rm F8}\Bigr)\\ 
\langle 8,3/2;SSP|H|8,3/2;SSA\rangle' & = & -\frac{3}{2\sqrt{6}}{\rm F7} 
\end{eqnarray}
\noindent{$\underline{J^P=1/2^-:N^*/\Lambda^*/\Sigma^*}$}
\begin{eqnarray}
'\langle 1,1/2;SSP|H|1,1/2;SSP\rangle' & = & {\rm F5} + {\rm F6} + {\rm 
F13} \hspace{2cm}
\label{eq:PTHOOFTA} \\
\langle 8,1/2;SSP|H|8,1/2;SSP\rangle & =  & \frac{1}{4}{\rm F6} + {\rm F11} 
- {\rm F12} + \frac{1}{4}{\rm F13} 
\label{eq:PTHOOFTB} \\
'\langle 8,1/2;SSP|H|8,1/2;SSP\rangle' & = & \frac{1}{4}\Bigl(4{\rm F5} + 
{\rm F6} + {\rm F13}\Bigr)
\label{eq:PTHOOFTC} \\  
\langle 8,1/2;SSA|H|8,1/2;SSA\rangle  & = & \frac{3}{2} \Bigl( 
-\frac{1}{\sqrt{3}}{\rm F2} + 
\frac{1}{2}{\rm F3}+\frac{1}{6}{\rm F4} + \frac{1}{3}{\rm F9} 
- \frac{1}{3}{\rm F10}  \Bigr) \\  
\langle 8,1/2;SSP|H|8,1/2;SSP\rangle' & =  & 
\frac{1}{2}\Bigl(-\frac{1}{2}{\rm F6} - {\rm F12} + \frac{1}{2}{\rm F13} 
\Bigr)\\
\langle 8,1/2;SSA|H|8,1/2;SSP\rangle & = & \frac{1}{2} \Bigl( 
-\frac{\sqrt{3}}{2}{\rm F7} + \frac{1}{2}{\rm F8} -2{\rm F14} 
- \frac{1}{2}{\rm F15} 
\Bigr)\\  
\langle 8,1/2;SSA|H|8,1/2;SSP\rangle' & = & \frac{\sqrt{3}}{4} 
\Bigl({\rm F7} - \frac{1}{\sqrt{3}}{\rm F8} 
+ \frac{1}{\sqrt{3}}{\rm F15} \Bigr)
\end{eqnarray}

The functions F1 to F15 are given by
\begin{eqnarray}
{\rm F1} & = & -\frac{8}{15}\int\!\! dr R\:r^2 \biggl(G_0F_3 
- G_3F_0\biggr)^2 \\
{\rm F2} & = & -\frac{2}{3\sqrt{3}}\int\!\! dr R\:r^2 
\Biggl[ 
\frac{8}{5}\biggl(G_0G_3F_0F_3\biggr) + 
\frac{3}{5}\biggl(G_0F_3+G_3F_0\biggr)^2 + \biggl(G_0G_3-F_0F_3\biggr)^2 
\Biggr] \\
{\rm F3} & = & \frac{2}{3}\int\!\! dr R\:r^2
\Biggl[
\biggl(G_0^2-F_0^2\biggr)\biggl(G_3^2-F_3^2\biggr) + \frac{4}{5}\biggl(
G_0F_0G_3F_3\biggr)
+ \frac{4}{5}\biggl(G_0F_3+G_3F_0\biggr)^2
\Biggr] \\
{\rm F4} & = & \frac{2}{3}\int\!\! dr R\:r^2
\Biggl[
\biggl(G_0^2-F_0^2\biggr)\biggl(G_3^2-F_3^2\biggr) 
- \frac{4}{15}\biggl(G_0F_0G_3F_3\biggr)
 \nonumber \\
&     &\hspace{5cm} - \frac{2}{3}\biggl(G_0G_3-F_0F_3\biggr)^2
+ \frac{1}{5}\biggl(G_0F_3+G_3F_0\biggr)^2
\Biggr] \\
{\rm F5} & = & \int\!\! dr R\:r^2 \biggl(G_0^2+F_0^2\biggr)^2 \\
{\rm F6} & = & 2\int\!\! dr R\:r^2 
\Biggl[
\biggl(G_0^2-F_0^2\biggr)\biggl(G_2^2-F_2^2\biggr) 
- \frac{4}{3}\biggl(G_0F_0G_2F_2\biggr) 
- \frac{1}{3}\biggl(G_0G_2-F_0F_2\biggr)^2 \nonumber \\
&     & \hspace{5cm} + \biggl(F_0G_2+G_0F_2\biggr)^2
\Biggr] \\
{\rm F7} & = & \frac{2\sqrt{2}}{3}\int\!\! dr R\:r^2
\biggl(F_0F_3+G_0G_3\biggr)\biggl(G_0G_2+F_0F_2\biggr)\\
{\rm F8} & = & \sqrt{\frac{8}{27}}\int\!\! dr R\:r^2
\Biggl[
-2\biggl(G_0F_0\biggr)\biggl(G_2F_3+G_3F_2\biggr) \nonumber \\
&     & \hspace{5cm} + \biggl(G_0G_3-F_0F_3\biggr)\biggl(G_0G_2-F_0F_2\biggr)
\Biggr] \\
{\rm F9} & = & \frac{2}{3}\int\!\! dr R\:r^2 
\Biggl[
\biggl(G_0^2-F_0^2\biggr)\biggl(G_3^2-F_3^2\biggr) 
+ \frac{4}{15}\biggl(G_0G_3F_0F_3\biggr)
\nonumber \\
&     & \hspace{5cm} -\frac{1}{3}\biggl(G_0G_3-F_0F_3\biggr)^2
 + \frac{3}{5}(G_0F_3+G_3F_0\biggr)^2 
\Biggr] \\
{\rm F10} & = & \frac{2}{3} \int\!\! dr R\:r^2 
\Biggl[
-\frac{16}{15}\biggl(G_0F_0G_3F_3\biggr) 
- \frac{2}{3}\biggl(G_0G_3-F_0F_3\biggr)^2
+ \frac{2}{5}\biggl(G_0F_3+G_3F_0\biggr)^2 
\Biggr] \\
{\rm F11} & = & 2 \int\!\! dr Rr^2 
\Biggl[
\biggl(G_0^2-F_0^2\biggr)\biggl(G_2^2-F_2^2\biggr) 
+ \frac{4}{3}\biggl(G_0F_0G_2F_2\biggr)
- \frac{2}{3}\biggl(G_0G_2-F_0F_2\biggr)^2
\Biggr] \\
{\rm F12} & = & 2\int\!\! dr Rr^2
\Biggl[
-\frac{8}{3}\biggl(G_0F_0G_2F_2\biggr) 
+ \frac{1}{3}\biggl(G_0G_2-F_0F_2\biggr)^2
+\biggl(F_0G_2+G_0F_2\biggr)^2
\Biggr] \\
{\rm F13} & = & 2\int\!\! dr R\:r^2
\Biggl[
\frac{1}{3}\biggl(G_0G_2+F_0F_2\biggr)^2-\biggl(G_0F_2-G_2F_0\biggr)^2
\Biggr] \\
{\rm F14} & = & \sqrt{\frac{8}{27}}\int\!\! dr R\:r^2
\Biggl[
2\biggl(G_0F_0\biggr)\biggl(G_2F_3+G_3F_2\biggr) \nonumber \\
&    & \hspace{5cm} 
+ \biggl(G_0G_3-F_0F_3\biggr)\biggl(G_0G_2-F_0F_2\biggr)
\Biggr] \\
{\rm F15} & = & -\sqrt{\frac{32}{27}}\int\!\! dr R\:r^2
\biggl(G_0F_0\biggr)\biggl(G_2F_3+G_3F_2\biggr) 
\end{eqnarray}
The radial functions $G_0(r), 
F_0(r), G_2(r), F_2(r), G_3(r)$ and $F_3(r)$ are 
defined in Eqs.~(\ref{eq:SWAVE}) 
to (\ref{eq:AWAVE}) and are shown in Figure~1 for $m_q$ = 7.5 and 300 MeV.
\section{Fourier transformed meson--quark source function}
\label{appendixd}
Definitions of operators ${\cal O}^{a, l}_{MN}(\hat{q})$ and 
corresponding functions $F^l(|\vec{q}\:|)$ appearing in the Fourier 
transformed meson--quark source function, 
Eq.~(\ref{eq:VFUNCTION}), are presented in this Appendix. 
The functions $F^l(|\vec{q}\:|)$ in 
Eq.~(\ref{eq:VFUNCTION}) are defined as
\begin{eqnarray}
F_{SS}^{l=1}(|\vec{q}\:|)   & = &
(-i 8\pi) \int dr r^2 j_1(|\vec{q}\:|r) \frac{g_\phi}{(g_\chi \chi(r))^2}G_0(r)F_0(r) \\
F_{PP}^{l=1}(|\vec{q}\:|)   & = &
(-i 8\pi) \int dr r^2 j_1(|\vec{q}\:|r) \frac{g_\phi}{(g_\chi \chi(r))^2}G_2(r)F_2(r) \\
F_{AA}^{l=1}(|\vec{q}\:|) & = &
(-i 8\pi) \int dr r^2 j_1(|\vec{q}\:|r) \frac{g_\phi}{(g_\chi \chi(r))^2}G_3(r)F_3(r) \\
F_{AP}^{l=1}(|\vec{q}\:|) & = &
(-i 4\pi) \int dr r^2 j_1(|\vec{q}\:|r) \frac{g_\phi}{(g_\chi \chi(r))^2}
\Big( G_2(r)F_3(r) + G_3(r)F_3(r) \Big) \\
F_{AA}^{l=3}(|\vec{q}\:|) & = &
(+i 8\pi) \int dr r^2 j_3(|\vec{q}\:|r) \frac{g_\phi}{(g_\chi 
\chi(r))^2}G_3(r)F_3(r), 
\end{eqnarray}
where $j_l(|\vec{q}\:|r)$ is the usual spherical Bessel function of order $l$. 
The corresponding single quark operators ${\cal O}^{a, l}_{MN}(\hat{q})$ are
\begin{eqnarray}
{\cal O}^{a, l=1}_{SS}(\hat{q}) & = &
\lambda^a \left( \chi_m^{1/2} \right)^\dagger 
\hat{q} \cdot \vec{\sigma} 
 \left( \chi_m^{1/2} \right) \\
{\cal O}^{a, l=1}_{PP}(\hat{q}) & = &
\lambda^a \left( \chi_m^{1/2} \right)^\dagger
\hat{q} \cdot \vec{\sigma} 
 \left( \chi_m^{1/2} \right)\\
{\cal O}^{a, l=1}_{AA}(\hat{q}) & = &
\lambda^a \frac{1}{5} \left( \chi_m^{3/2} \right)^\dagger
\sigma_b^{[3/2,1/2]} \hat{q} \cdot \vec{\sigma}  \sigma_b^{[1/2,3/2]} 
\left( \chi_m^{3/2} \right) \\
{\cal O}^{a, l=1}_{AP}(\hat{q}) & = &
\lambda^a \left( \chi_m^{3/2} \right)^\dagger
\hat{q} \cdot \vec{\sigma}^{[3/2,1/2]} 
\left( \chi_m^{1/2} \right)\\
{\cal O}^{a, l=3}_{AA}(\hat{q}) & = &
\lambda^a \left( \chi_m^{3/2} \right)^\dagger 
\Bigg[ \hat{q} \cdot \vec{\sigma}^{[3/2,1/2]} \hat{q} \cdot \vec{\sigma}
\hat{q} \cdot \vec{\sigma}^{[1/2,3/2]}
- \frac{1}{5} \sigma_b^{[3/2,1/2]} \hat{q} \cdot \vec{\sigma}  
\sigma_b^{[1/2,3/2]} \Bigg]
\left( \chi_m^{3/2} \right)
\end{eqnarray}
In the above definitions $\chi_{m}^{j}$ is a $2j+1$ component 
spinor analogous to 
the two component Pauli spinors for $j$=1/2 case. The operators $\sigma_a$
with $a = 1, 2, 3$ are the usual Pauli matrices, while 
$\sigma_a^{[1/2,3/2]}$ are 
$(2 \times 4)$ Multipole Transition Matrices given by \cite{eeg83},
\begin{eqnarray}
\sigma_1^{[1/2,3/2]} & = & \frac{1}{\sqrt{6}}
\left( 
\begin{array}{cccc}
-\sqrt{3} & 0  & 1 & 0 \\
0            & -1 & 0 & \sqrt{3} 
\end{array}
\right), \\
\sigma_2^{[1/2,3/2]} & = & \frac{-i}{\sqrt{6}}
\left( 
\begin{array}{cccc}
\sqrt{3} & 0  & 1 & 0 \\
0           & 1  & 0 & \sqrt{3} 
\end{array}
\right), \\
\sigma_3^{[1/2,3/2]} & = & \sqrt{\frac{2}{3}}
\left( 
\begin{array}{cccc}
0 & 1  & 0 & 0 \\
0 & 0  & 1 & 0 
\end{array}
\right),
\label{eq:MTM}
\end{eqnarray}
with the property,
\begin{equation}
\left( \sigma^{[1/2,3/2]}_a \right)^\dagger = 
\sigma^{[3/2,1/2]}_a. 
\end{equation}

\vfill\eject

\centerline{\bf TABLE CAPTIONS}
\vskip 1cm
TABLE~I. \hspace{0.1 cm} 
Eigenvalues $\epsilon_i$ of $H^{(0)}_{\rm Toy}$ as defined in 
Eq~(\ref{eq:EIGEN}) for $i = S, P$ and $A$ quark spinor states with masses 
$m_q$ = 7.5 and 300 MeV.
\vskip 0.75cm
TABLE~II. \hspace{0.1 cm} 
Values of scalar charge ${\cal S}_i $ defined in 
Eq.~(\ref{eq:SCALAR}) for $i= S, P$ and $A$ with quark masses $m_q=7.5$ MeV 
and 300 MeV.
\vskip 0.75cm
TABLE~III. \hspace{0.1 cm} 
Well--established low--lying negative parity baryon states 
considered in this work \cite{pdg94}. 
Spin--parity assignments and experimental mass ranges 
of the resonances are shown in columns two and three, respectively. In column 
four the $L-S$ coupled basis with the notation 
$|70,^{2S+1}SU(3)_J\rangle$ allowed by the projection 
prescription discussed in Section~\ref{diagonal} are shown for each state.
\vskip 0.75cm
TABLE~IV. \hspace{0.1 cm} 
Matrix elements of  $H^{(0)}_{\rm Toy}$, Eq.~(\ref{H0Toy}), and 
$H_{\rm 't~Hooft}$, 
Eq.~(\ref{eq:Hooft}), in MeV for negative parity baryons.
These matrix elements are 
obtained using $m_s = m_{u,d}$ = 7.5 MeV and using the 
projection prescription 
discussed in Section~\ref{diagonal}. The values of model parameters used in
the calculation are given in Section~\ref{ccm}. The fourth column shows the 
matrix elements 
of OGE interaction obtained in the MIT bag model 
evaluated in the SU(3) limit with vanishing current quark masses. 
The OGE matrix elements 
are multiplied by the strong coupling constant $\alpha_s$.
\vskip 0.75cm
TABLE~V. \hspace{0.1 cm} 
Bare masses and SU(6) spin--flavor wave functions of negative parity 
baryons $N^*$, $\Delta^*$, $\Lambda^*$ and $\Sigma^*$ in the CCM. 
Masses and wave functions obtained by using 
the toy model Hamiltonian, $H^{(0)}_{\rm Toy}$, alone
are shown in the second column while the entries in the third column 
are calculated using the bare CCM Hamiltonian, $H_{\rm Bare} =  
H^{(0)}_{\rm toy} + H_{\rm 't~Hooft}$. 
Baryon masses in MeV are indicated within parenthesis and the three--quark 
wave functions are expressed in the $L-S$ coupled basis. 
Column four shows the bare masses of
hyperons in MeV after being corrected for $m_s - m_{u,d}$ mass difference 
as explained in Section~\ref{toymodel}. 
\vskip 0.75cm
TABLE~6. \hspace{0.1 cm} 
Relative percentages of spin--flavor contents in the $L-S$ coupled basis
for low--lying negative parity $J^P=3/2^-$ and $1/2^-$ $\Lambda^*$ hyperons 
in the CCM of this work, the MIT bag model with massless quarks and the 
NRQM \cite{isg77}. 
\vfill\eject
%
%
\begin{table}
\begin{tabular}{|c||c|c|c|}
$m_q$ (MeV) & $\epsilon_S$ (MeV) & $\epsilon_P$ (MeV) & $\epsilon_A$ (MeV) \\
\hline\hline
7.5  & 340 & 550 & 490 \\
\hline
300 & 595 & 772 & 743 \\
\end{tabular}
\vskip 0.5cm\caption{ }
\label{H0energies}
\end{table}
%
%
\begin{table}
\begin{tabular}{|c||c|c|c|}
$m_q$ (MeV) & ${\cal S}_S$ & ${\cal S}_P$ & ${\cal S}_A$ \\
\hline
7.5&0.61&0.33&0.54\\
\hline
300&0.79&0.63&0.71\\
\end{tabular}
\vskip 0.5cm\caption{ }
\label{stable}
\end{table}
%
%
\begin{table}
\begin{tabular}{|c||c|c|c|}
Resonance & $J^P$ & Mass Range (MeV) & $|70,^{2S+1}SU(3)_J\rangle$ \\
	\hline \hline
$N(1675)$ & $5/2^- $ & 1670 -- 1685 & $|^48_{5/2}\rangle$  \\
	\hline
$N(1700)$ & $3/2^-$ & 1650 -- 1750 & $|^48_{3/2}\rangle\;\; 
|^28_{3/2}\rangle$ \\
	\hline
$N(1520)$ & $3/2^- $ & 1515 -- 1530 & $|^48_{3/2}\rangle\;\; 
|^28_{3/2}\rangle$  \\
	\hline
$N(1650)$ & $1/2^- $ & 1640 -- 1680 & $|^48_{1/2}\rangle\;\; 
|^28_{1/2}\rangle$  \\
	\hline
$N(1535)$ & $1/2^- $ & 1520 -- 1555 &  $|^48_{1/2}\rangle\;\; 
|^28_{1/2}\rangle$ \\
	\hline
$\Delta(1700)$ &$3/2^- $& 1670 -- 1770 &$|^210_{3/2}\rangle$  \\
	\hline
$\Delta(1620)$ &$1/2^-$ & 1615 -- 1675 &$|^210_{1/2}\rangle$  \\
	\hline
$\Lambda(1830)$ &$5/2^-$ & 1810 -- 1830 &$|^48_{5/2}\rangle$  \\
	\hline
$\Lambda(1690)$  &$3/2^-$ & 1685 -- 1695 &$|^48_{3/2}\rangle\;\; 
|^28_{3/2}\rangle\;\; |^21_{3/2}\rangle$  \\
	\hline
$\Lambda(1520)$  &$3/2^-$ & 1518 -- 1520 &$|^48_{3/2}\rangle\;\; 
|^28_{3/2}\rangle\;\; |^21_{3/2}\rangle$ \\
	\hline
$\Lambda(1800)$  &$1/2^-$ & 1720 -- 1850 &$|^48_{1/2}\rangle\;\; 
|^28_{1/2}\rangle\;\; |^21_{1/2}\rangle$ \\
	\hline
$\Lambda(1670)$  &$1/2^-$ & 1660 -- 1680 &$|^48_{1/2}\rangle\;\; 
|^28_{1/2}\rangle\;\; |^21_{1/2}\rangle$  \\
	\hline
$\Lambda(1405)$  &$1/2^-$ & 1400 -- 1410 & $|^48_{1/2}\rangle\;\; 
|^28_{1/2}\rangle\;\; |^21_{1/2}\rangle$ \\
	\hline
$\Sigma(1775)$ &$5/2^-$& 1770 -- 1780 &$|^48_{5/2}\rangle$   \\
	\hline
$\Sigma(1940)$ &$3/2^-$ & 1900 -- 1950 &$|^48_{3/2}\rangle\;\; 
|^28_{3/2}\rangle\;\; |^210_{3/2}\rangle$   \\
	\hline
$\Sigma(1670)$ &$3/2^-$& 1665 --1685 &$|^48_{3/2}\rangle\;\; 
|^28_{3/2}\rangle\;\; |^210_{3/2}\rangle$   \\
	\hline
$\Sigma(1750)$ &$1/2^-$ & 1730 -- 1800 &$|^48_{1/2}\rangle\;\; 
|^28_{1/2}\rangle\;\; |^210_{1/2}\rangle$ \\   
\end{tabular}
\vskip 0.5cm\caption{ }
\label{States}
\end{table}
\vfill\eject
%
%
\begin{table}
\begin{tabular}{|c||c|c|c|}
State & $H^{(0)}_{\rm Toy}$ & $H_{\rm 't~Hooft}$  & $H_{\rm OGE}$ \\
	\hline\hline
$N(3/2^-)$ &
$\left( 
\begin{array}{cc}
1503 & 21\\
21 & 1483
\end{array}
\right)$ & 
$\left( 
\begin{array}{cc}
-31 & 0.2\\
0.2 & -31
\end{array}
\right)$ & 
$\alpha_s \left( 
\begin{array}{cc}
60 & -1\\
-1 & 47
\end{array}
\right)$\\
\hline
$N(1/2^-)$ &
$\left( 
\begin{array}{cc}
1523 & 13 \\
13 & 1503
\end{array}
\right)$ & 
$\left( 
\begin{array}{cc}
178 & 89 \\
89 & -12
\end{array}
\right)$& 
$\alpha_s \left( 
\begin{array}{cc}
-23 & -15\\
-15 & -24
\end{array}
\right)$ \\
\hline
$\Lambda(3/2^-)$ &
$\left( 
\begin{array}{ccc}
1470 & 0 & 0\\
0 & 1503 & 21\\
0 & 21 & 1483
\end{array}
\right)$ & 
$\left( 
\begin{array}{ccc}
-90 & 0 & 0\\
0 & -31 & 0.2\\
0 & 0.2 & -31
\end{array}
\right)$ & 
$\alpha_s \left( 
\begin{array}{ccc}
-12 & 0 & 0\\
0 & 60 & -1\\
0 & -1 & 47
\end{array}
\right)$ \\
\hline
$\Lambda(1/2^-)$ &
$\left( 
\begin{array}{ccc}
1530 & 0 & 0 \\
0 & 1523 & 13\\
0 & 13 & 1503
\end{array}
\right)$ & 
$\left( 
\begin{array}{ccc}
-10 & 0 & 0\\
0 & 178 &  89\\
0 & 89 & -12
\end{array}
\right)$ & 
$\alpha_s \left( 
\begin{array}{ccc}
-61 & 0 & 0\\
0 & -23 & -15\\
0 & -15 & -24
\end{array}
\right)$\\
\hline
$\Sigma(3/2^-)$ &
$\left( 
\begin{array}{ccc}
1497 & 0 & 0 \\
0 & 1503 & 21\\
0 & 21 & 1483 \\
\end{array}
\right)$ & 
$\left( 
\begin{array}{ccc}
0 & 0 & 0\\
0 & -31 & 0.2\\
0 & 0.2 & -31
\end{array}
\right)$ & 
$\alpha_s \left( 
\begin{array}{ccc}
85 & 0 & 0 \\
0 & 60 & -1 \\
0 & -1 & 47 
\end{array}
\right)$\\
\hline
$\Sigma(1/2^-)$ &
$\left( 
\begin{array}{ccc}
1477 & 0 & 0\\
0 & 1523 & 13\\
0 & 13 & 1503
\end{array}
\right)$ & 
$\left( 
\begin{array}{ccc}
0 & 0 & 0\\
0 & 178 & 89\\
0 & 89 & -12
\end{array}
\right)$ & 
$\alpha_s \left( 
\begin{array}{ccc}
55 & 0 & 0\\
0 & -23 & -15\\
0 & -15 & -24
\end{array}
\right)$\\
\end{tabular}
\vskip 0.5cm\caption{ }
\label{Melements}
\end{table}
\vfill\eject
%
%
\begin{table}
\begin{tabular}{|c||c|c|c|}
State & $H^{(0)}_{\rm Toy}$ & $H_{\rm Bare} =  H^{(0)}_{\rm Toy} + 
H_{\rm 't~Hooft}$ & $\Delta s$ \\
	\hline\hline
$N(5/2^-)$ & $|(1470) \rangle = |^48_{5/2}\rangle$ & 
$|(1470) \rangle = |^48_{5/2}\rangle$ & N/A\\
	\hline
$N(3/2^-)_1$ & $|(1518)\rangle = 0.823|^48_{3/2}\rangle + 0.567|^28_{3/2}
\rangle$ & 
$|(1485) \rangle = 0.846|^48_{3/2}\rangle + 0.534|^28_{3/2}\rangle$ & N/A \\
	\hline
$N(3/2^-)_2$ & $|(1473)\rangle = -0.567|^48_{3/2}\rangle + 0.823|^28_{3/2}
\rangle$ &
$|(1439)\rangle = -0.534|^48_{3/2}\rangle + 0.846|^28_{3/2}\rangle$  & N/A \\
	\hline
$N(1/2^-)_1$ & $|(1529)\rangle = 0.897|^48_{1/2}\rangle + 0.442|^28_{1/2}
\rangle$ &
$|(1742)\rangle = 0.927|^48_{1/2}\rangle + 0.376|^28_{1/2}\rangle$ & N/A \\
	\hline
$N(1/2^-)_2$ & $|(1496)\rangle = -0.442|^48_{1/2}\rangle + 0.897|^28_{1/2}
\rangle $ &
$|(1450)\rangle = -0.376|^48_{1/2}\rangle + 0.927|^28_{1/2}\rangle$& N/A \\
	\hline\hline
$\Delta(3/2^-)$ & $|(1497)\rangle = |^210_{3/2}\rangle$ & 
$|(1497)\rangle = |^210_{3/2}\rangle$ & N/A \\
	\hline
$\Delta(1/2^-)$ & $|(1477)\rangle = |^210_{1/2}\rangle$ & 
$|(1477)\rangle |^210_{1/2}\rangle$ & N/A \\
	\hline\hline
$\Lambda(5/2^-)$ & $|(1470)\rangle = |^48_{5/2}\rangle$  
& $|(1470)\rangle = |^48_{5/2}\rangle$ & 1654 \\
	\hline
$\Lambda(3/2^-)_1$ & $|(1518)\rangle = 0.823|^48_{3/2}\rangle 
+ 0.567|^28_{3/2}\rangle$ & 
$|(1485) \rangle = 0.846|^48_{3/2}\rangle + 0.534|^28_{3/2}\rangle$ & 1667\\
	\hline
$\Lambda(3/2^-)_2$ & $|(1473)\rangle = -0.567|^48_{3/2}\rangle 
+ 0.823|^28_{3/2}\rangle$ &
$|(1439)\rangle = -0.534|^48_{3/2}\rangle + 0.846|^28_{3/2}\rangle$ & 1618 \\
	\hline
$\Lambda(3/2^-)_3$ & $|(1470)\rangle = |^21_{3/2}\rangle$ & 
$|(1380)\rangle = |^21_{3/2}\rangle$ & 1557 \\
	\hline
$\Lambda(1/2^-)_1$ & $|(1529)\rangle = 0.897|^48_{1/2}\rangle 
+ 0.442|^28_{1/2}\rangle$ 
& $|(1742)\rangle = 0.927|^48_{1/2}\rangle + 0.376|^28_{1/2}\rangle$  
& 1925 \\
	\hline
$\Lambda(1/2^-)_2$ & $|(1530)\rangle = |^21_{1/2}\rangle$ 
& $|(1520)\rangle = |^21_{1/2}\rangle$ & 1676 \\
	\hline
$\Lambda(1/2^-)_3$ & $|(1496)\rangle = -0.442|^48_{1/2}\rangle 
+ 0.897|^28_{1/2}\rangle$ 
& $|(1450)\rangle = -0.376|^48_{1/2}\rangle + 0.927|^28_{1/2}\rangle$ & 
1628   \\
	\hline\hline
$\Sigma(5/2^-)$ & $|(1470)\rangle = |^48_{5/2}\rangle$ 
& $|(1470)\rangle = |^48_{5/2}\rangle$ & 1640   \\
	\hline
$\Sigma(3/2^-)_1$ & $|(1497)\rangle = |^410_{3/2}\rangle$ 
& $|(1497)\rangle = |^410_{3/2}\rangle$ & 1665   \\
	\hline
$\Sigma(3/2^-)_2$ & $|(1518)\rangle = 0.823|^48_{3/2}\rangle 
+ 0.567|^28_{3/2}\rangle$ & 
$|(1485) \rangle = 0.846|^48_{3/2}\rangle + 0.534|^28_{3/2}\rangle$ & 1638\\
	\hline
$\Sigma(3/2^-)_3$ & $|(1473)\rangle = -0.567|^48_{3/2}\rangle 
+ 0.823|^28_{3/2}\rangle$ &
$|(1439)\rangle = -0.534|^48_{3/2}\rangle + 0.846|^28_{3/2}\rangle$  & 1601\\
	\hline
$\Sigma(1/2^-)_1$ & $|(1529)\rangle = 0.897|^48_{1/2}\rangle 
+ 0.442|^28_{1/2}\rangle$ 
& $|(1742)\rangle = 0.927|^48_{1/2}\rangle + 0.376|^28_{1/2}\rangle$ 
& 1881   \\
	\hline
$\Sigma(1/2^-)_2$ & $|(1477)\rangle = |^210_{1/2}\rangle$ 
& $|(1477)\rangle = |^210_{1/2}\rangle$ & 1652  \\
	\hline
$\Sigma(1/2^-)_3$ & $|(1496)\rangle = -0.442|^48_{1/2}\rangle 
+ 0.897|^28_{1/2}\rangle$ 
& $|(1450)\rangle = -0.376|^48_{1/2}\rangle + 0.927|^28_{1/2}\rangle$ 
& 1619  \\
\end{tabular}
\vskip 0.5cm\caption{ }
\label{Results}
\end{table}
%

%
\begin{table}
\begin{tabular}{|c||c||c|c|c|}          
Model ($\rightarrow$) & \multicolumn{4}{c|}{Chiral Confining Model} \\ \hline
Hyperon state ($\downarrow $)   & Bare Mass (MeV) & $|^{2}1_J\rangle$ (\%) & 
$|^{4}8_J\rangle$ (\%) & $|^{2}8_J\rangle$ (\%) \\ \hline\hline   
$\Lambda(3/2^{-})_{1}$ & 1667 &  0.0  & 71.6     & 28.4      \\ \hline   
$\Lambda(3/2^{-})_{2}$ & 1618 &  0.0  & 28.4     & 71.6      \\ \hline
$\Lambda(3/2^{-})_{3}$ & 1557 & 100  & 0.0       & 0.0     \\ \hline
$\Lambda(1/2^{-})_{1}$ & 1925 & 0.0   & 85.9     & 14.1       \\ \hline   
$\Lambda(1/2^{-})_{2}$ & 1676 & 100  & 0.0       & 0.0      \\ \hline
$\Lambda(1/2^{-})_{3}$ & 1628 & 0.0   & 14.1     & 85.9       \\ \hline
Model ($\rightarrow$) & \multicolumn{4}{c|}{MIT Bag Model} \\ \hline
Hyperon state ($\downarrow $)   & Mass (MeV) & $|^{2}1_J\rangle$  (\%) & 
$|^{4}8_J\rangle$ (\%)  & $|^{2}8_J\rangle$ (\%) \\ \hline\hline   
$\Lambda(3/2^{-})_{1}$ & 1705 & 0.0  & 80.8 & 19.2  \\ \hline   
$\Lambda(3/2^{-})_{2}$ & 1620 & 0.0  & 19.2 &  80.8  \\ \hline
$\Lambda(3/2^{-})_{3}$ & 1526 & 100 & 0.0   & 0.0  \\ \hline
$\Lambda(1/2^{-})_{1}$ & 1598 & 0.0  & 100  & 0.0     \\ \hline   
$\Lambda(1/2^{-})_{2}$ & 1563 & 0.0  & 0.0   & 100 \\ \hline
$\Lambda(1/2^{-})_{3}$ & 1552 & 100 & 0.0   & 0.0  \\ \hline
Model ($\rightarrow$) & \multicolumn{4}{c|}{Non--Relativistic Quark Model} 
\\ \hline
Hyperon state ($\downarrow $)   & Physical Mass (MeV) & $|^{2}1_J\rangle$ 
(\%) & 
$|^{4}8_J\rangle$ (\%) & $|^{2}8_J\rangle$ (\%) \\ \hline\hline   
$\Lambda(3/2^{-})_{1}$ & 1880 & 0.2  & 98.0  & 1.2     \\ \hline   
$\Lambda(3/2^{-})_{2}$ & 1690 & 16.0 & 1.4   & 82.8     \\ \hline
$\Lambda(3/2^{-})_{3}$ & 1490 & 82.8 & 1.0
   & 16.0     \\ \hline
$\Lambda(1/2^{-})_{1}$ & 1800 & 3.2   & 72.3 & 25.0     \\ \hline   
$\Lambda(1/2^{-})_{2}$ & 1650 & 10.0   & 33.6 & 56.3     \\ \hline
$\Lambda(1/2^{-})_{3}$ & 1490 & 81.0 & 0.4   & 18.5      \\ 
\end{tabular}                                                       
\vskip 0.5cm\caption{ }
\label{Percent}
\end{table}
\vfill\eject
\centerline{FIGURE CAPTIONS}
\vskip 1cm
FIG~1. \hspace{0.1 cm} The upper, $G(r)$, and the
lower, $F(r)$, radial functions for quark spinors in states $S_{1/2}$, 
$P_{1/2}$ and 
$P_{3/2}$ as defined in Eqs.~(\ref{eq:SWAVE}) to (\ref{eq:AWAVE}). 
The solid and dashed lines indicate radial functions obtained with 
quark masses
of $m_q$ =7.5 MeV and $m_q$ = 300 MeV, respecitively. 
a) $S_{1/2}$ radial functions $G_0(r)$ and $F_0(r)$. The radial 
profile of the 
$\chi^{(0)}$ field is also shown for comaprison using the dot--dashed line.
b) $P_{1/2}$ radial functions $G_2(r)$ and $F_2(r)$.
c) $P_{3/2}$ radial functions $G_3(r)$ and $F_3(r)$. 
Note that the distance 
shown in the horizontal axis is given in mesic units of $1/\mu$.
\vskip 0.75cm
FIG~2. \hspace{0.1 cm} Normalized $N N \pi$ dipole form factor 
$R(N N \pi) \equiv 
G_{N N \pi}^{l=1}(|\vec{q}\:|)/G_{N N \pi}^{l=1}(0)$ for 
a) $N=N(5/2^-)$, 
b) $N=N(3/2^-)_1$ (dashed line), $N=N(3/2^-)_2$ (solid line) and 
c) $N=N(1/2^-)_1$ (dashed line), $N=N(1/2^-)_2$ (solid line).
\vskip 0.75cm
FIG~3. \hspace{0.1 cm} Normalized $\Lambda_1 N k$ dipole form factor
$R(\Lambda_1 N k) \equiv
G_{\Lambda_1 N k}^{l=1}(|\vec{q}\:|)/G_{\Lambda_1 N k}^{l=1}(0)$ for
a) $N=N(3/2^-)_1$ (dashed line), $N=N(3/2^-)_2$ (solid line) and
b) $N=N(1/2^-)_1$ and $N=N(1/2^-)_2$ (solid line). Note that
$R(\Lambda_1 N k)$ involving $N(1/2^-)$ states are identical.
\vfill\eject
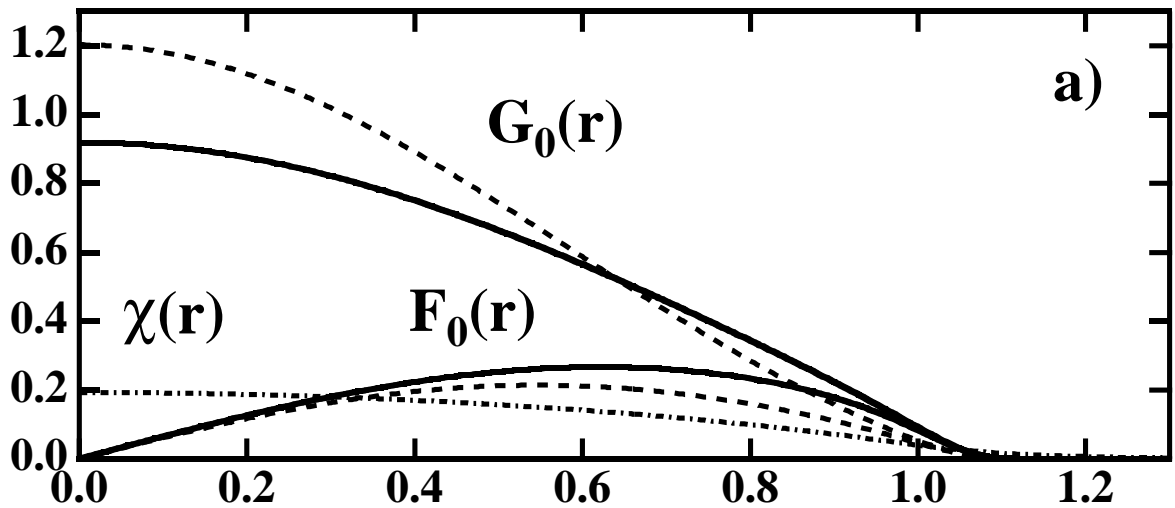
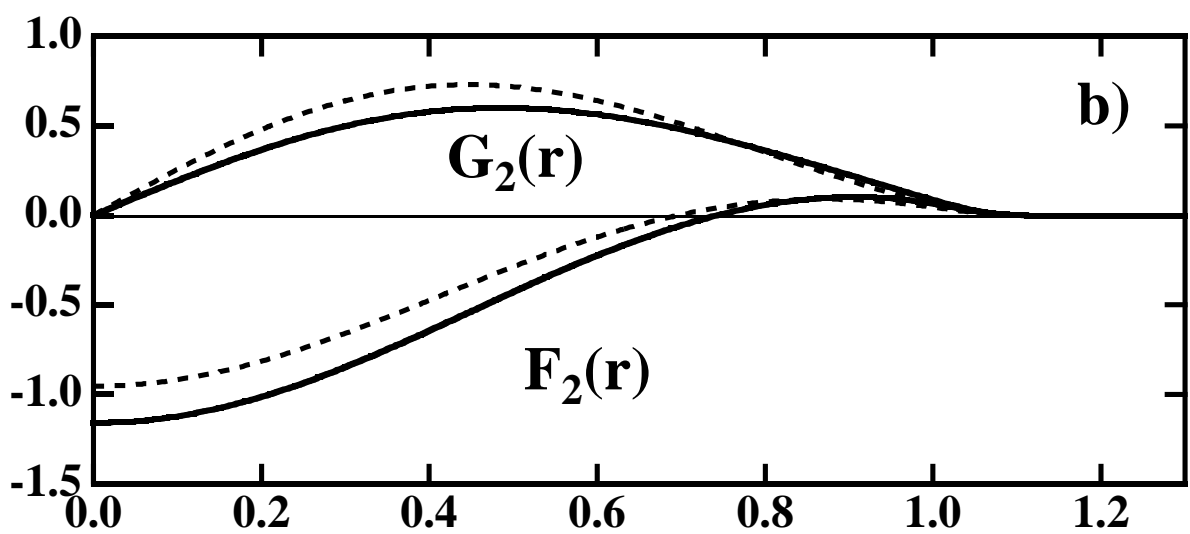
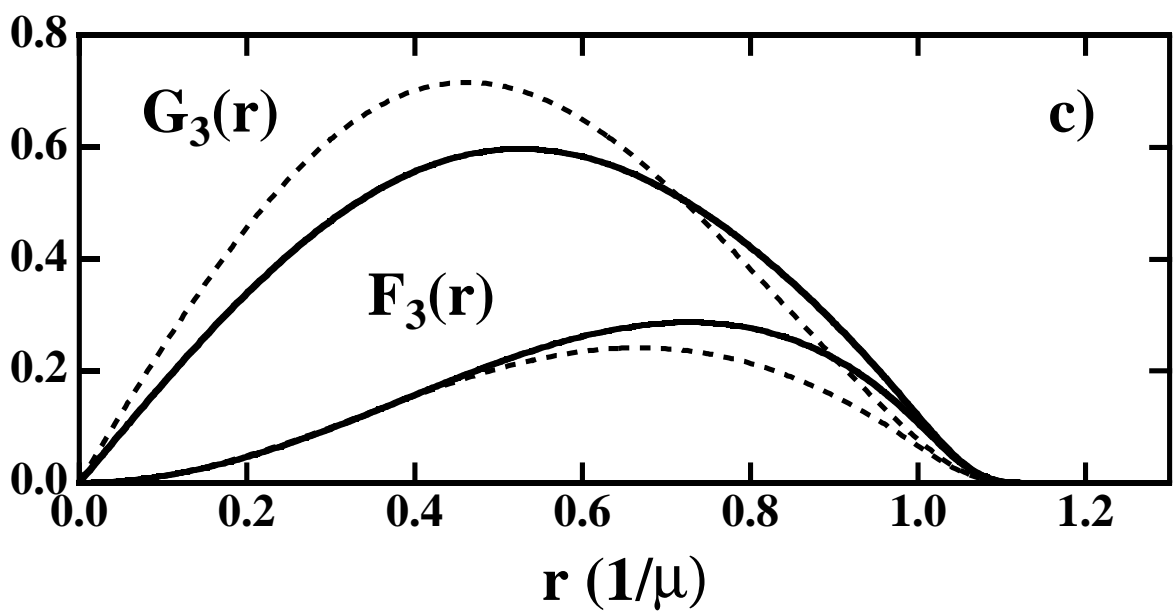
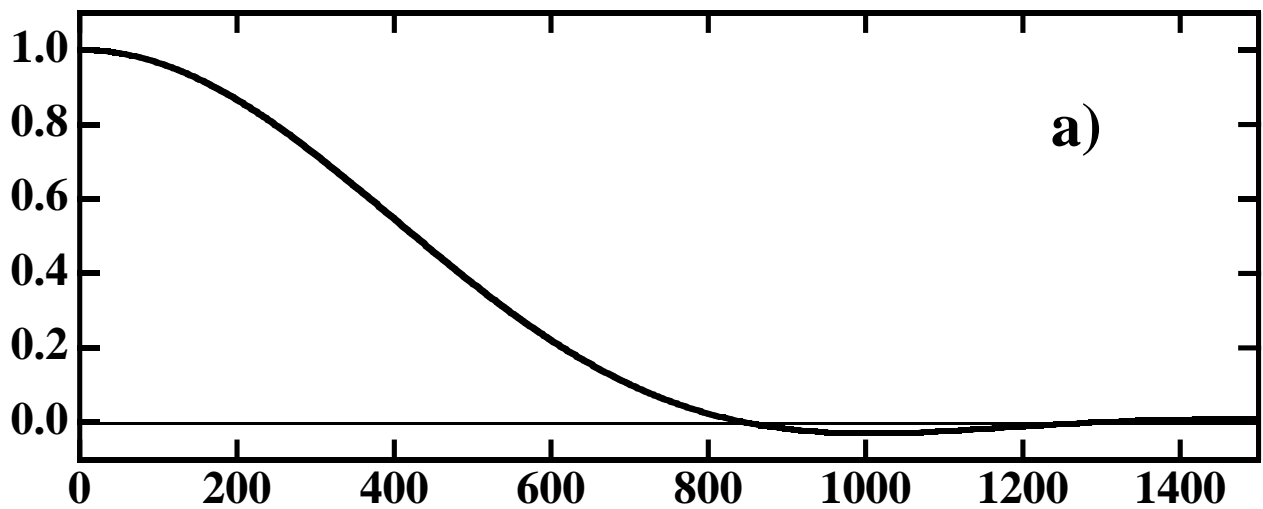
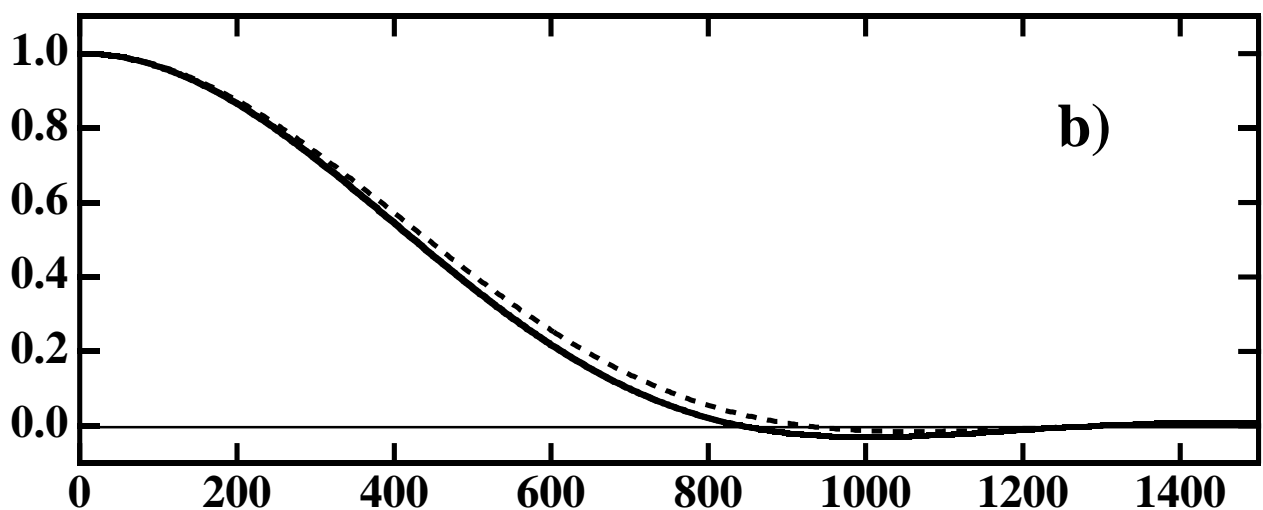
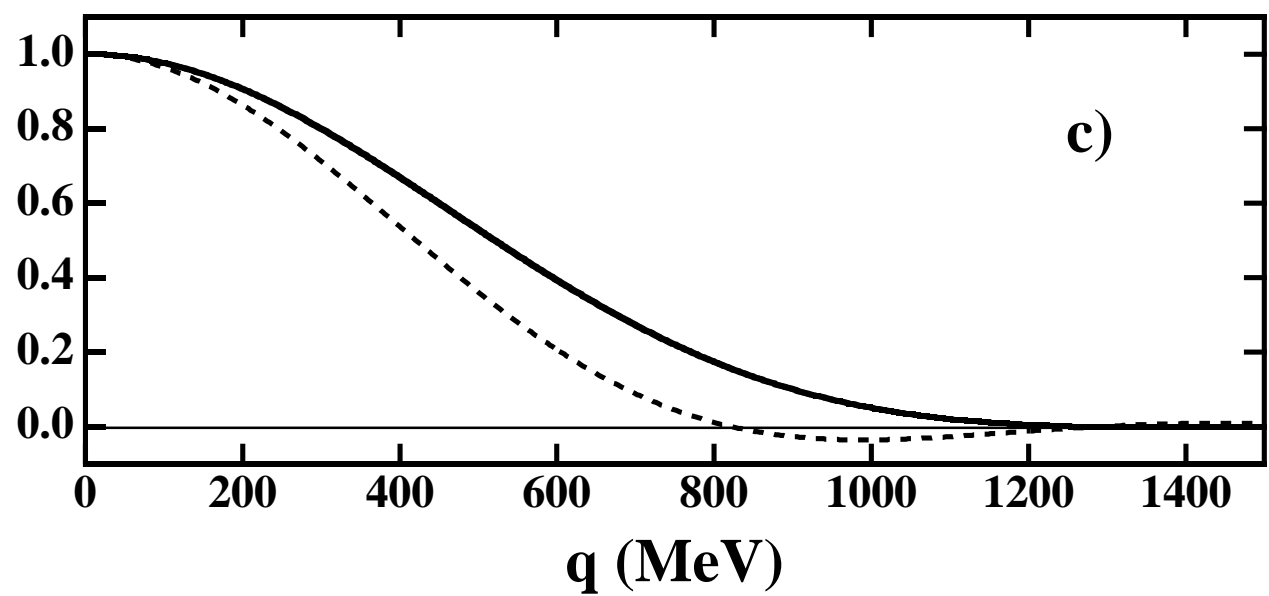
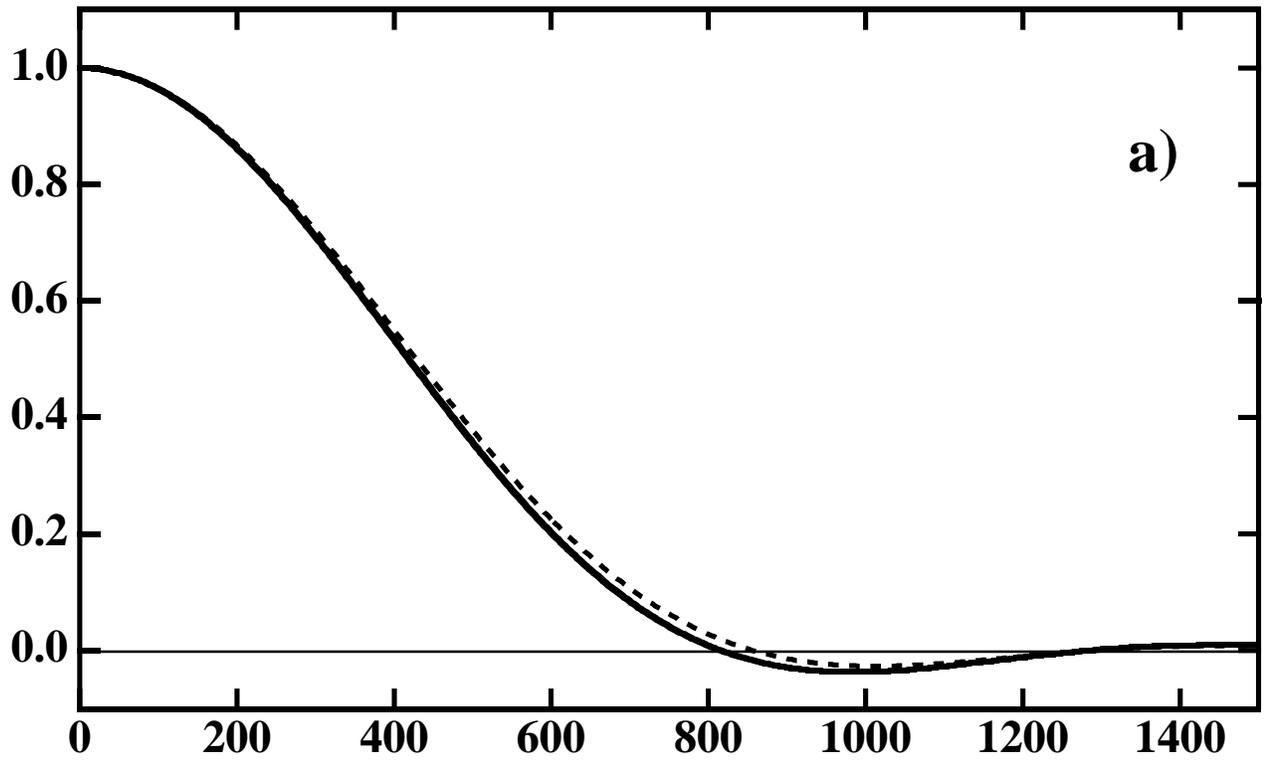
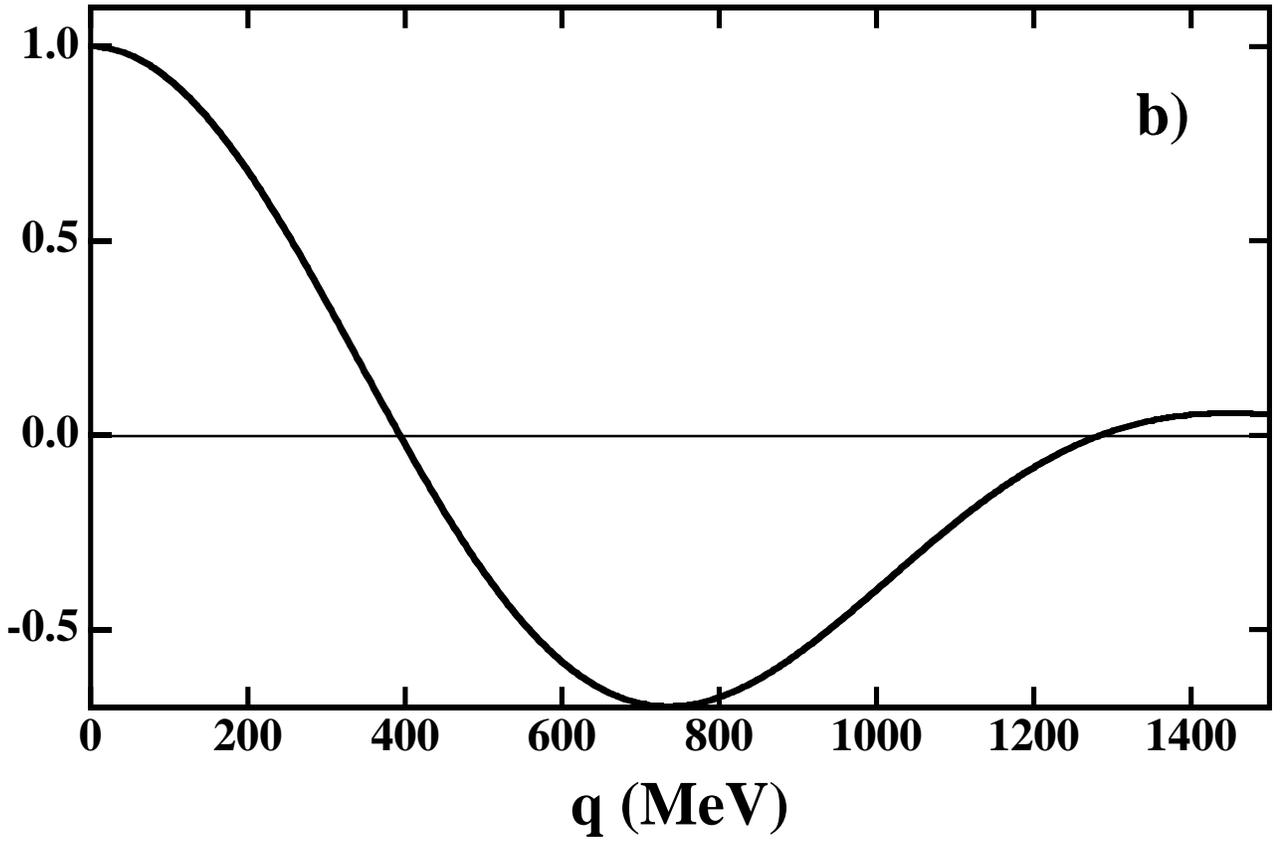
\end{document}